\documentclass[slac_one]{revtex4}
\usepackage{graphicx}
\usepackage{color}
\usepackage{fancyhdr}
\usepackage{rotating}

\newcommand{\RR}{{\rm R}}

\newcommand{\SLASH}[2]{\makebox[#2ex][l]{$#1$}/}
\newcommand{\Eslash}{\SLASH{E}{.5}\,}
\begin{document}

%%%%%%%%%%%%%%%%%%%%%%%%%%%%%%%%%%%%%%%%%%%%%%%%%%

\title{{\small{2005 ALCPG \& ILC Workshops - Snowmass,
U.S.A.}}\\ %% Please keep this conference title here
\vspace{12pt}
ILC: Physics Scenarios}

\author{W.~Kilian and P.~M.~Zerwas}
\affiliation{Deutsches Elektronen-Synchrotron DESY,
   D-22603 Hamburg, Germany}

\begin{abstract}
Experiments in the energy range from the scale of electroweak symmetry
breaking to the TeV scale are expected to be crucial for unraveling
the microscopic structure of matter and forces. The high precision
which should be achieved in experiments at lepton colliders, 
is a necessary ingredient for providing a comprehensive
picture of the mechanism breaking the electroweak symmetries and
generating mass, the unification of forces, involving most likely
supersymmetry, and the structure of space-time at small distances. In
addition, clarifying the nature of the particles which build up 
cold dark matter in the universe, needs a lepton collider to match 
the high experimental precision which will be reached in cosmology 
experiments.
\centerline{%
\unitlength1mm
\begin{picture}(0,60)
\put(-13,3){\includegraphics[height=5cm]{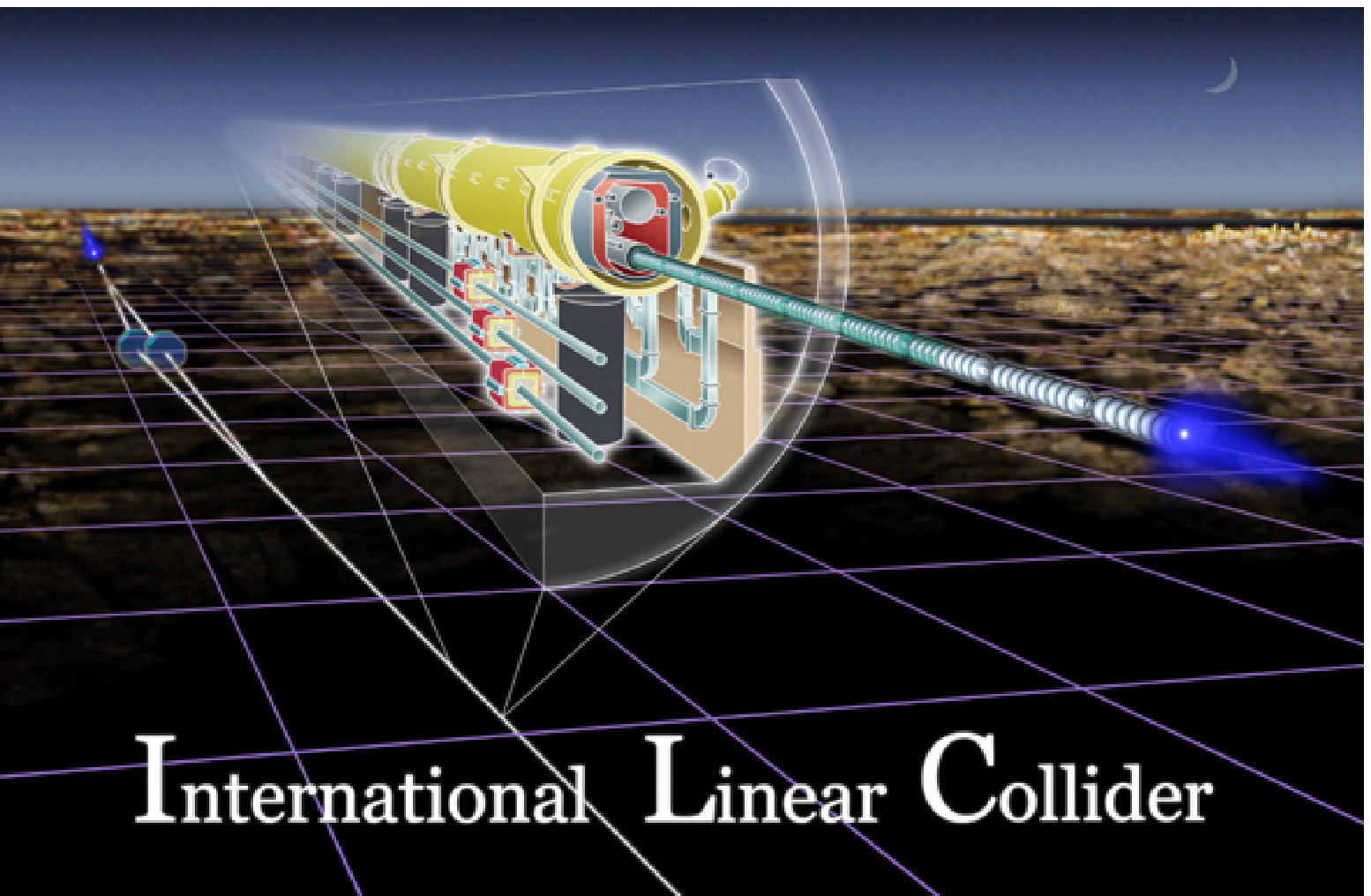}}
\end{picture}
}
\end{abstract}

\maketitle

\thispagestyle{fancy}
%%%%%%%%%%%%%%%%%%%%%%%%%%%%%%%%%%%%%%%%%%%%%%%%%%%%%%%%%%%%%%%%%%%%%%%%%%%%%%%%%%%

\section{INTRODUCTION}

High-energy physics has been tremendously successful in unraveling the
basic laws of nature in the microcosm. With the Standard Model of
particle physics a picture has emerged which adequately describes 
the structure of matter and forces. However, this picture is still 
incomplete internally, and
externally, driven by theoretical arguments and experimental
observations, the model should be embedded in a more comprehensive
theory which unifies the different degrees of freedom. These points
lead us in a natural way to a set of crucial experimental
questions. Answering these questions will unify our view of the
microscopic world and thus deepen our understanding of the universe
enormously. 

Derived from our present knowledge of particle physics, solutions
to the following problems, which are central to physics in general,
must be approached experimentally: 
\begin{itemize}
\item[--] 
  the mechanism responsible for breaking the electroweak symmetries
  and generating mass;
\item[--]
  the unification of forces, including gravity in the end; 
\item[--]
  the structure of space-time at small distances.  
\end{itemize}
This set of fundamental problems is complemented by a new branch in the development
of particle physics: 
\begin{itemize}
\item[--]
  the connection to cosmology. 
\end{itemize}
Besides the nature of particles which may form components of cold dark
matter in the universe, several other problems connect microscopic
physics and cosmology, the baryon asymmetry in the universe being a
prominent example. 

Based on the present picture of physics, the scientific value of any
new accelerator is determined by the unique contributions the facility
can offer in approaching solutions to these problems. 

We are in the fortunate position that the next generation of
accelerators holds the promise of providing answers to these questions
indeed. They will greatly advance the understanding of the microscopic
world in particular and the universe as a whole. With the Large Hadron
Collider LHC, to be completed at CERN in a couple of years, a first
decisive step will be taken. From this machine which will operate at
the TeV energy frontier, we expect breakthrough discoveries in the
complex of electroweak symmetry breaking and in the physics area
beyond the Standard Model. However, this hadron facility must be
complemented with a lepton collider which will play a key role in
drawing a comprehensive and high-resolution picture of electroweak
symmetry breaking and of the physics
layer
 underneath the Standard Model. Our present knowledge of physics
is expected to converge to a unified picture in this layer. 

The $e^+e^-$ Linear Collider ILC, which is now in the design phase,
would be the counterpart in the tandem with LHC, cf.\
Refs.~\cite{r1.1,r1.2,r1.3}. In analogy to the relation between LEP
and Tevatron, the ILC energy of 1 TeV in the lepton sector is equivalent
in many aspects to the higher LHC energy, effectively about 5 TeV in the
quark sector. Moreover, by including the characteristic scale of
electroweak symmetry breaking, the ILC covers one of the most crucial
energy ranges in particle physics. Discoveries at LHC may also point
to physics scales beyond the reach of ILC; this area could be accessed
later by a multi-TeV $e^+e^-$ facility \cite{r1.4}.

\subsection{Physics Scenarios}

\noindent{\underline{\it Electroweak Symmetry Breaking and Higgs Mechanism}} 

\vspace{0.5\baselineskip}
The mechanism which breaks the electroweak symmetries, is the
still missing cornerstone of the Standard Model. High-precision 
analyses strongly suggest the Higgs mechanism, including a light Higgs boson, 
to be responsible  for the breaking of the electroweak symmetries and for
generating the masses of the fundamental particles \cite{r1.4.1}. If the Higgs 
boson will be discovered at LHC, it must be established experimentally that 
this mechanism is indeed responsible for generating the masses of the 
particles. The precision with which this question can be answered 
at ILC, exceeds the LHC by an order of magnitude. In addition, 
in the most probable light mass range ILC provides the unique
opportunity for establishing the Higgs self-energy potential, which
is the essential {\it agens} for inducing the symmetry breaking. 

In extensions of the Standard Model, like supersymmetric theories or
Little Higgs theories, the Higgs sector is much more complex. A
spectrum of Higgs particles will in general be realized, demanding
precision studies of masses, mixing and couplings to explore the
structure of the Higgs sector. 

If the standard Higgs mechanism, including a set of Higgs particles,
were not realized in nature, but alternatively a higgs-less theory as
suggested, for example, in theories of electroweak symmetry breaking
by new strong interactions at low scales, cf. Ref. \cite{r1.4.2}, 
such a scenario could be
explored in the scattering of electroweak bosons at LHC and
ILC. However, taking advantage of the less complex final-state
topology at the lepton collider ILC, experiments at this machine can
cover the entire threshold region of the new strong interactions and
open the door to an arena of novel interactions. Other higgs-less
scenarios, as formulated in some theories of extra space dimensions,
also give rise to new interactions between the standard electroweak
gauge bosons mediated by new TeV scale resonances.

\vspace{\baselineskip}
\noindent{\underline{\it Unification and Supersymmetry}} 

\vspace{0.5\baselineskip}
Progress in particle physics has opened the path to the truly unified
understanding of nature. The unification of the electromagnetic, weak
and strong interactions is strongly indicated by the evolution of the
couplings merging at high energies, cf. Refs. \cite{r1.5} ,
and expected to be joined by gravity in the ultimate unification 
near the Planck scale. A key role in the evolution is played by supersymmetry,
cf. Ref. \cite{r1.6}. LHC has the potential to discover supersymmetry in the 
next few years, and the theoretical concept can be verified in conjunction 
with ILC which is an essential instrument in this process.  

  Supersymmetry embraces several of the fundamental points introduced
at the beginning -- providing a stable bridge between the scale of
electroweak symmetry breaking and the Planck scale; leading to the
unification of the standard couplings and paving the path for
including gravity in particle physics. In addition, the lightest
supersymmetric particle is a compelling candidate for forming a
component of the large amount of cold dark matter observed in the
universe. Thus, this theory could not only play a fundamental role in
particle physics but also links particle physics closely with
cosmology.  

   In fact, high-precision measurements of electroweak observables,
combined with constraints from the observation of the cold dark matter
density by WMAP, allow for a large area of fairly low-scale
supersymmetry parameters, though no firm conclusions can be drawn as
yet. In the favorable case a significant fraction of the non-colored
supersymmetric particles, i.e., partners of the photon, of the
electron etc, should be observed at ILC operating in the first phase
at 500 GeV, and more in the upgraded 1 TeV phase of the machine. LHC
would play the complementary role for colored particles, the
supersymmetric partners of the quarks and gluons.  

   Quite generally, apart from exceptional corners of parameter space,
LHC experiments will discover supersymmetric particles if this
symmetry is realized in nature not far above the electroweak
scale. However, the spectrum of particles in this new world that can
be detected at LHC will remain incomplete, particularly in the light
non-colored sector. Moreover, the precision in the determination of
their properties, like masses, mixings and couplings, remains
limited. Operating ILC will, first, lead to a comprehensive view of
the spectrum of light particles and, second, improve the accuracy in
measuring their properties by one to two orders of magnitude.

   Both points are very important for several reasons. Foremost, the
completeness of the spectrum and the greatly improved accuracies will
allow us to extrapolate the parameters to the unification scale where
the fundamental supersymmetric theory and the microscopic picture of
its breaking mechanism can be reconstructed.  

   This way we can study the structure of physics at scales close to
the Planck scale. This provides us with the unique opportunity to shed
light on an energy domain where the roots of particle physics in
particular, and physics in general, may be located. Information on
this area from other branches of particle physics, potentially proton
decay experiments etc, will remain very scarce so that the telescope
character of high-precision high-energy experiments, in coherent
LHC+ILC analyses, is of exceptionally high value.  

   High precision is also required in exploring the properties of the
lightest supersymmetric particle which may contribute to the observed 
density of cold dark matter in the universe. Anticipating improved results 
from cold dark matter measurements in the near future, the accuracy of
a lepton collider will be needed for masses, mixings and couplings to match 
eventually the accuracy of cosmology data. In addition, once the particle 
properties are determined accurately, observed fluxes in astroparticle
search experiments can be exploited to map the distribution of cold
dark matter in the universe. Thus ILC experimental results could reach
far beyond the domain of particle physics.

\vspace{\baselineskip}
\noindent{\underline{\it Extra Space Dimensions}} 

\vspace{0.5\baselineskip}
If extra space dimensions in the universe, cf. 
Refs. \cite{r1.7,r1.8,r1.9}, are realized already at
low energies, the experimental determination of the fundamental scale
of gravity and the number of dimensions are of central interest.
Starting these analyses with LHC, the picture can be refined
considerably at ILC. By varying the energy of the collider, these two
characteristics of gravity and space-time at short distances can be
disentangled. By observing masses and widths of excited graviton
states in other scenarios, the length scale and the curvature in an
additional fifth dimension of space-time can be determined.  

  Many other measurements could be performed in this area, e.g.,
measurements of the spin of gravity fields, mixings of scalar fields
etc, so that a large set of observables could be exploited at ILC
which, joined with LHC results, would enable us to zoom in on the
underlying theoretical picture.

\subsection{Basic Experimental Parameters}

It is generally assumed that the International Linear Collider ILC
will be operated in two phases. In the first phase the cm energy will
reach $\sqrt{s} =$ 500 GeV, in the second phase 1 TeV. In each of the
phases a total integrated luminosity of 1 ab$^{-1}$ is expected to be
accumulated when the runs are completed.  The first phase gives access
to light Higgs bosons, the top quark, light supersymmetric particles,
the second phase to strong electroweak symmetry breaking, heavy new
particles in the Higgs and supersymmetric sectors, extra space
dimensions and other high-scale phenomena. Some scenarios may suggest
extensions of the linear collider program beyond the TeV energy.  

Experiments at ILC will focus on high-precision analyses. If the
electron and positron beams are polarized, typically $P_{e^-} \sim
90\%$ and $P_{e^+} \sim 60\%$, the experimental potential of the
machine can truly be exhausted, cf.\ Ref.~\cite{r2}. In addition to
longitudinally polarized beams, spin rotators can generate
transversely polarized electron/positron beams. The polarization of
the electron beam is a necessary condition for many experimental
analyses while the polarization of the positron beam is generally
viewed as an auxiliary tool which however may turn out to be crucial 
in some special physics scenarios.  

The luminosity in running the machine as an $e^-e^-$ collider is
significantly smaller as the electrons repel each other when the
bunches of the two colliding beams traverse each other.  

In addition to the high-energy electron-positron collider mode, the
machine can be operated in the GigaZ mode. Running at low energies on
top of the $Z$-boson resonance, some 1 billion events, i.e., a factor
fifty more than at LEP, may be collected within a few months. Combined
with $W$ and $top$ threshold analyses, this leads to the ultimate precision 
in the electroweak sector in the foreseeable future. Both electron and
positron polarization is essential for these analyses.  

Finally, by means of Compton back-scattering of laser light, a
fraction of 80\% of the incoming electron/positron energy can be
transferred to the final-state photon, cf.\ Ref.~\cite{r3}.  The
spectrum is maximal at the upper edge if the incoming $e^-/e^+$ beam
and the laser photon beam are longitudinally polarized with opposite
helicities. In this way colliding $e \gamma$ and $\gamma \gamma$
experiments can be performed with 90\% and 80\% of the total $e^+e^-$
energy, respectively, and about one third of the luminosity
accumulating in a 20\% margin below the maximum possible energy. In
some scenarios these modes open up unique discovery channels for
particles, in the Higgs and slepton sectors of supersymmetric
theories, or in the particle towers of compositeness models, for
example. 

%%%%%%%%%%%%%%%%%%%%%%%%%%%%%%%%%%%%%%%%%%%%%%%%%%%%%%%%%%%%%%%%%%%%%%%%%%%%%%%%%%%%%%%%%%%%

\section{ELECTROWEAK SYMMETRY BREAKING}                               

Unraveling the mechanism which breaks the electroweak symmetries and
generates the masses of the fundamental standard particles ---
electroweak gauge bosons, leptons and quarks --- is one of the key
problems of particle physics, cf.\ Refs.~\cite{r4.1,r4.2}.  Theoretical
realizations span a wide range of scenarios extending from weak to
strong breaking mechanisms.  Examples on one side are the Standard
Model and its supersymmetric extension involving light fundamental
Higgs fields, and new strong interaction models without a fundamental
Higgs field on the other side. Little-Higgs models are located in the
transition zone.  Symmetry breaking by specific boundary conditions
for gauge fields in the compactification of extra space dimensions
gives rise to higgs-less models. The forthcoming experiments at LHC
will lead to a breakthrough in revealing the breaking mechanism and in
making the first steps into the new territory while ILC should provide
the comprehensive understanding of the theory underlying the breaking
of the electroweak symmetries and the generation of mass. Thus the
experimental solution of this problem at LHC and ILC will unravel one
of the fundamental laws of nature.

\subsection{Higgs Mechanism in the Standard Model}
  
The analysis of the precision electroweak data from LEP, SLC and
elsewhere points clearly to a light mass value of the Higgs particle 
\cite{r5.1},
if the electroweak symmetries are broken by the Higgs mechanism in the
framework of the Standard Model:
\begin{equation}
  M_H = 91^{+45}_{-32} \;{\rm GeV} \,\,\, 
  {\rm and} \,\,\, M_H < 186 \;{\rm GeV} \;\; 
                                              (95\% \,{\rm CL}).
\end{equation}     
The direct search for the SM Higgs boson at LEP has set a lower limit
of 114 GeV on the Higgs mass \cite{r5.2}.  

The Higgs particle of the Standard Model is guaranteed to be
discovered at LHC, cf.\ Ref.~\cite{r6}. The combination of several
channels in different mass ranges gives rise to a large significance
for the detection, i.e., $> 5 \sigma$ for an integrated luminosity
of 30 fb$^{-1}$.  

After the discovery of the Higgs particle, it must be established
experimentally that the Higgs mechanism is responsible indeed for
breaking the electroweak symmetries and for generating the masses of
the fundamental particles. This requires the precise determination of
the profile of the Higgs particle. First steps in model-independent
analyses of its properties can be taken at LHC by performing precision
measurements of the Higgs mass, the ratios of some of the Higgs
couplings, and bounds on couplings \cite{r7}.  

\begin{figure}[h]
\begin{center}
\vskip 3mm
\includegraphics[height=8cm]{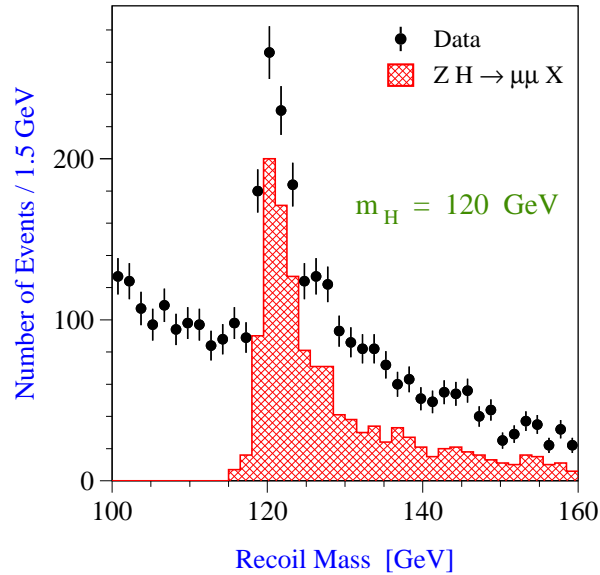}
\caption{Signal and background of inclusive Higgs boson production in
         Higgs-strahlung; Ref.~\cite{r8}.}
\label{F1}
\end{center}
\end{figure}

At ILC a clean sample of Higgs events can be generated in
Higgs-strahlung, $e^+e^- \to ZH$, and $WW$ fusion, $e^+e^- \to
\bar{\nu} \nu H$. The clear signals above small backgrounds, cf.\
Fig.~\ref{F1}, allow the model-independent high-precision determination
%%%%%%%%%%%%%%%%%
%%%%%%%%%%%%%%%%%
of the Higgs profile, besides the mass, the spin of the particle and,
most important, its couplings, including the trilinear self-coupling
in double-Higgs production. This information will be extracted from a
set of production cross sections and angular distributions, and from
decay branching ratios. Below a Higgs mass of 140 GeV a rich ensemble
of final states can be studied; the ensemble of channels is reduced for 
heavier Higgs masses.

\vspace{\baselineskip}
\noindent\underline{\it Higgs Couplings}  

\begin{figure}[h]
\begin{center}
\vskip 3mm
\includegraphics[height=8cm]{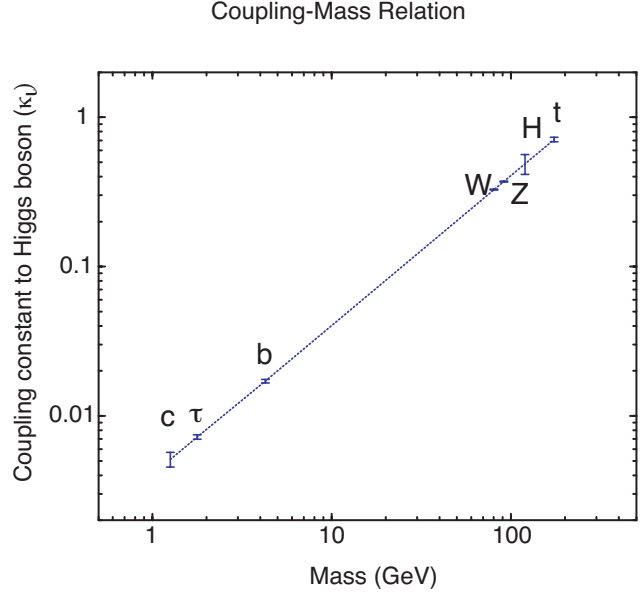}
\caption{The relation between the Higgs coupling of a particle and its
         mass in the Standard Model; Ref.~\cite{r10}.  The error bars
         correspond to the accuracy expected from ILC data.}
\label{F2}
\end{center}
\end{figure}

\vspace{0.5\baselineskip}
If the masses of the fundamental particles $p$ are generated by the
interaction with the Higgs field in the vacuum, the Higgs couplings
must grow with the particle masses:
\begin{equation}
g(Hpp) = (\sqrt{2} G_F)^{1/2} m_p.
\end{equation}
From the production cross sections for Higgs-strahlung and $WW$ fusion
the absolute values of the Higgs couplings to the electroweak gauge
bosons $Z$ and $W$ can be determined in a model-independent
way. Measuring the ratios of branching ratios involving quarks and
leptons on one side, and the electroweak gauge bosons on the other
side, also Higgs couplings to quarks and leptons can be determined in
a model-independent way. A special case is the Higgs-top coupling
which can be measured in Higgs radiation off top-quark pairs produced
in $e^+e^-$ annihilation.  The accuracy which can be achieved for
various couplings is predicted at the per-cent level \cite{r9}. How
well the Higgs coupling -- mass relation can be tested, is apparent
from Fig.~\ref{F2}
%%%%%%%%%%%%%%%%%
%%%%%%%%%%%%%%%%%
which clearly demonstrates the linear relation between the Higgs
couplings and the masses for typical particle species in the Standard
Model -- electroweak gauge bosons, quarks and leptons, up and down
types.

\vspace{\baselineskip}
\noindent\underline{\it Higgs Potential}  

\begin{figure}[htb]
\begin{center}
\includegraphics[height=1.7cm]{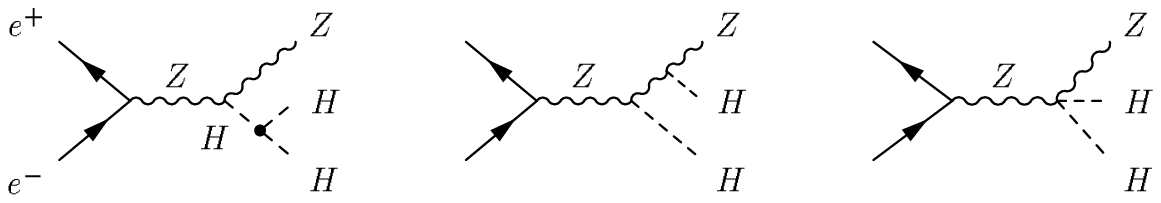}
\includegraphics[height=6cm]{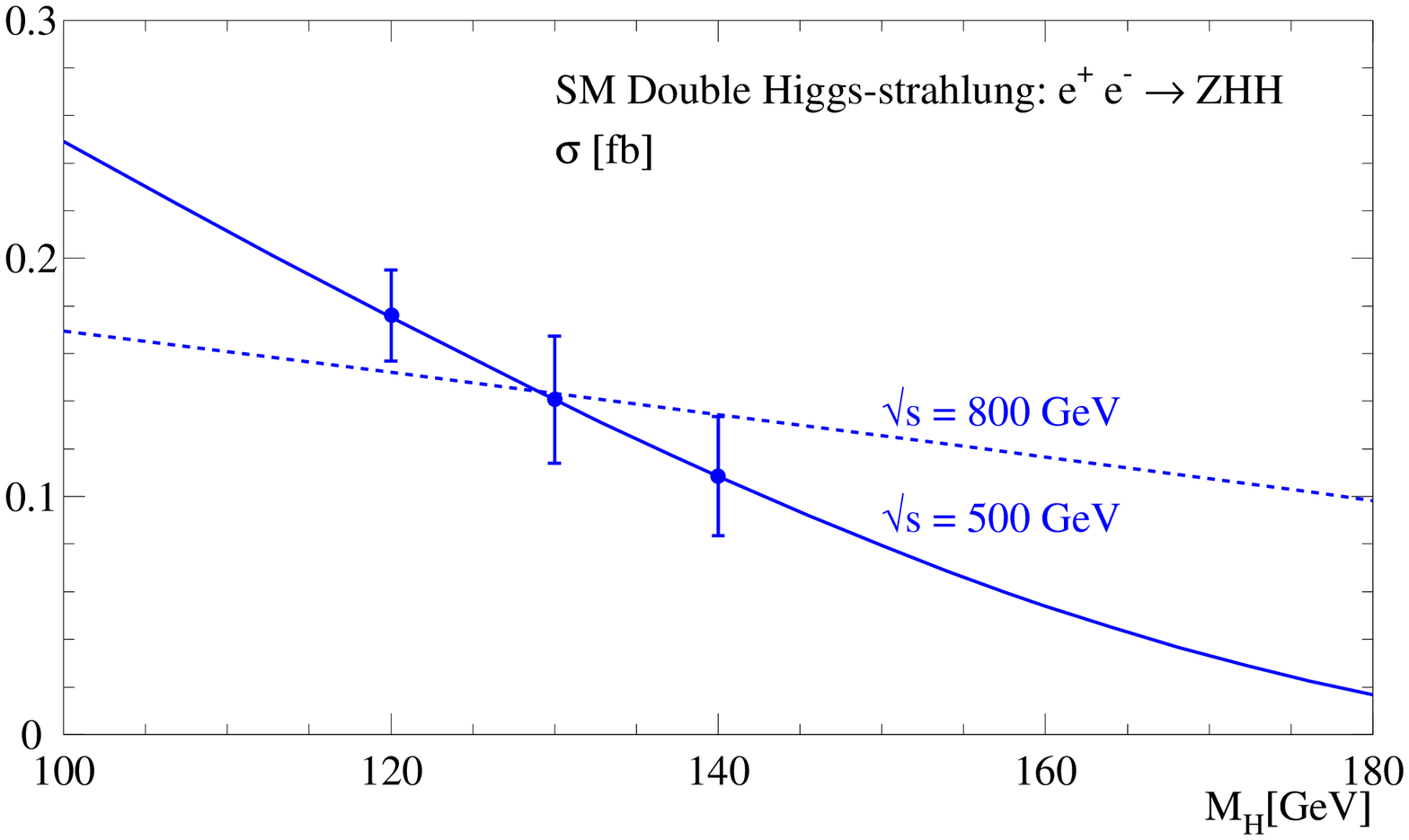}
\caption{Cross section of Higgs pair production for measurements of
         the triple Higgs coupling; Ref.~\cite{r11}.  The error bars
         correspond to the accuracy expected from ILC data.}
\label{F3}
\end{center}
\end{figure}

\nopagebreak
\vspace{0.5\baselineskip}
The specific form of the Higgs potential, $V \sim [|\phi|^2 -
v^2/2]^2$ shifts the ground state of the Higgs system to a non-zero
field strength, $v/\sqrt{2}$. Specifying the direction of the field
strength in charge space breaks the electroweak symmetries.  The gauge
and Yukawa interaction energy of other fields with the non-zero Higgs
field in the vacuum can be reinterpreted as the mass of these
particles. Expanding the potential about the minimum,
\begin{equation}
V=\frac{1}{2}M_H^2\,H^2 +\frac{1}{2}\frac{M_H^2}{v}\,H^3
                        + \frac{1}{8}\frac{M_H^2}{v^2}\,H^4
\end{equation}
the trilinear coupling plays the crucial role for the non-trivial
shape of the potential.  This parameter can be measured in the process
of double-Higgs production, $e^+e^- \to ZHH \;\; {\rm and} \;\;
\bar{\nu} \nu HH$, as exemplified in Fig.~\ref{F3}. The product of
small couplings and the large fraction of phase space absorbed by the
masses render the production cross sections small. Nevertheless, the
trilinear coupling is expected to be measured at ILC at a level of
15\% for Higgs masses below about 140 GeV. The less crucial quartic
coupling in the Standard Model seems out of reach for any collider in
the foreseeable future. Thus the element in the Higgs potential which
is most crucial for generating the Higgs medium in the ground state,
can be reconstructed at ILC. In the upper intermediate Higgs mass
range access to the trilinear coupling could be given by SLHC
\cite{r12}. 

%%%%%%%%%%%%%%%%%
%%%%%%%%%%%%%%%%%
                                                  
Direct measurements of the $ZZH$ coupling to an accuracy of 1\%, and
of the $HHH$ coupling to about 10\%, constrain scales of new physics
to about 3 and 1 TeV, respectively \cite{r12.1}. Since the microscopic 
dynamics of electroweak symmetry breaking is one of the central problems 
in particle physics, establishing these values would be a valuable
and unique result of experiments at ILC --- and even more so if
deviations from the SM predictions would be discovered.

\subsection{SUSY Higgs Bosons}

In supersymmetric theories the Higgs sector must be extended to at
least two iso-doublet fields so that five or more physical Higgs
particles are predicted. In the minimal extension the mass of the
lightest neutral scalar Higgs particle $h^0$ is bounded to about 140
GeV, while the masses of the heavy neutral scalar and pseudoscalar
Higgs bosons, $H^0$ and $A^0$, as well as the pair of charged Higgs
bosons, $H^\pm$, may range from the electroweak scale to the
(multi-)TeV region. The four heavy Higgs bosons tend to be nearly
mass-degenerate. The upper bound on the lightest Higgs mass is relaxed
to about 200 GeV in more general scenarios if the fields remain weakly
interacting up to the Planck scale as naturally assumed in
supersymmetric theories.  

\vspace{\baselineskip}
\noindent\underline{\it Minimal Supersymmetric Theory}  

\begin{figure}[h]
\begin{center}
\includegraphics[width=8cm]{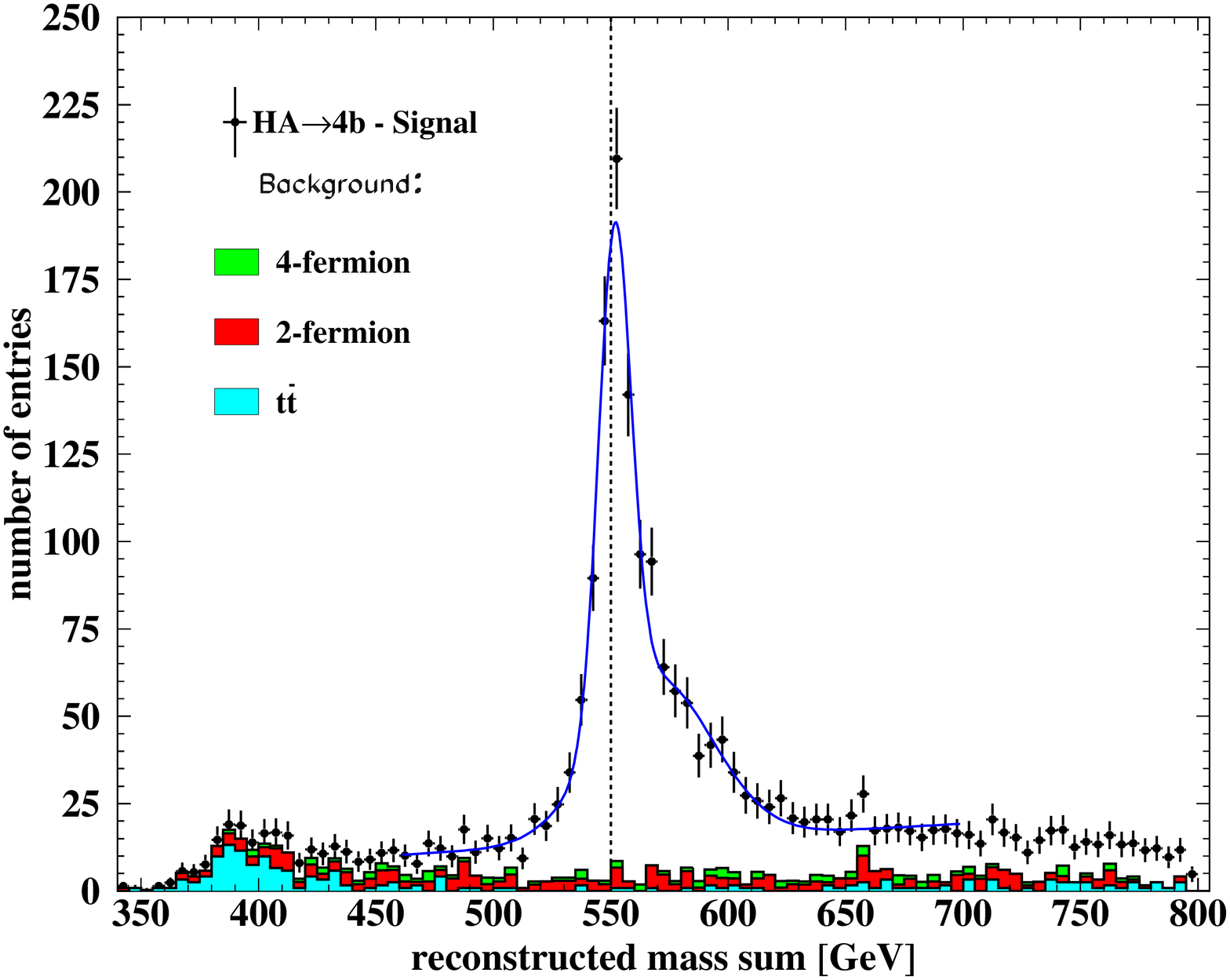}
\includegraphics[width=6.1cm]{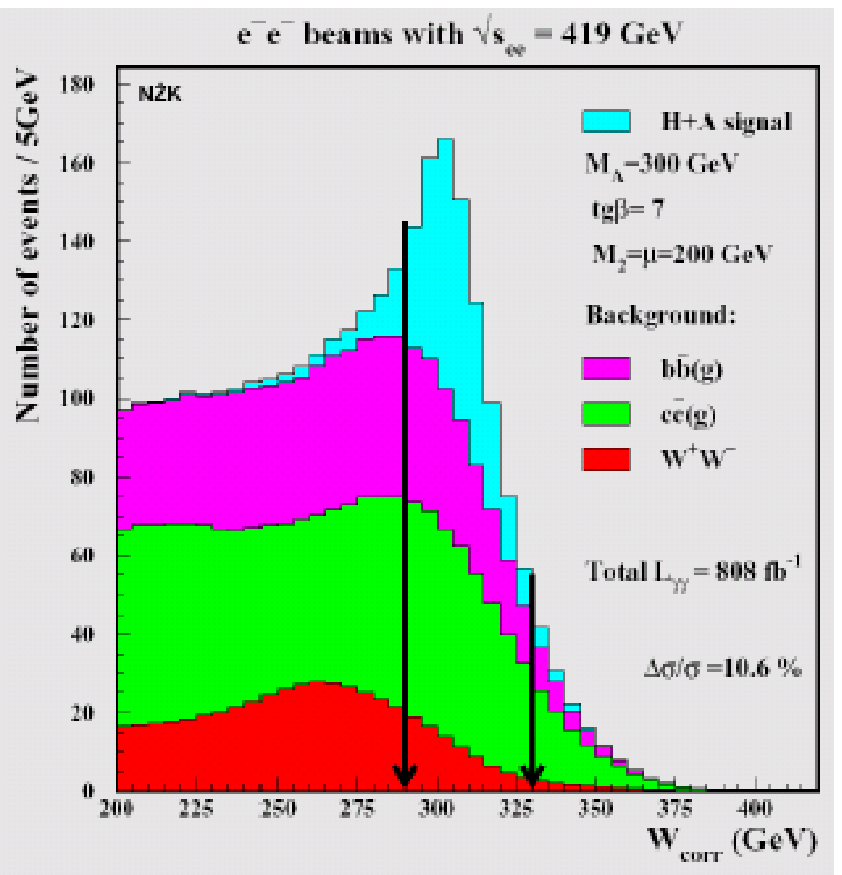}
\caption{Left: Pair production of MSSM Higgs bosons $HA$ in $e^+e^-$
         collisions; Ref.~\cite{r14}; Right: Single Higgs production
         $H$ and $A$ in $\gamma\gamma$ fusion at a photon collider;
         Ref.~\cite{r15}.}
\label{F4}
\end{center}
\end{figure}

\vspace{0.5\baselineskip}
While search and study of the light $h^0$ Higgs boson follows the
pattern summarized above for the SM Higgs boson in most of the
parameter space, the heavy scalar and pseudoscalar Higgs bosons are
produced in mixed pairs, in the same way as the charged Higgs bosons:
\begin{equation}
e^+e^- \to H^0 A^0 \;\; {\rm and} \;\; H^+H^- \, .
\end{equation}
For masses of the heavy Higgs bosons beyond about 200 GeV they cannot
be detected at LHC in a wedge in $M_A/\tan\beta$ parameter space that is
centered around the medium mixing angle $\tan\beta \sim 7$ and opens
up to high Higgs masses. The wedge can be covered by pair production
in $e^+e^-$ collisions for masses $M_{H,A} \leq \sqrt{s}/2$, i.e., up
to 500 GeV in the TeV phase of the machine. However, beyond this
range, single production in photon-photon collisions,
\begin{equation}
\gamma \gamma \to H^0 \;\; {\rm and} \;\; A^0
\end{equation}
can cover the wedge up to Higgs masses of 800 GeV if a fraction of
80\% of the total $e^+e^-$ energy is transferred to the $\gamma
\gamma$ system by Compton back-scattering of laser light
\cite{r13}. Thus, a $\gamma \gamma$ collider may be the only facility
in which, beyond the SM-type light Higgs boson, heavy Higgs bosons may
be discovered before a multi-TeV linear collider can be operated. It
is demonstrated in Fig.~\ref{F4} how well
%%%%%%%%%%%%%%%%%
%%%%%%%%%%%%%%%%%
the Higgs bosons can be detected in the two collider modes.  

High-precision measurements of the light Higgs mass can be exploited
to determine parameters in the theory which are difficult to measure
otherwise. By evaluating quantum corrections, the trilinear coupling
$A_t$, for example, may be calculated from
%%%%%%%%%%%%%%%%%
\begin{figure}[h]
\begin{center}
\includegraphics[width=6.6cm]{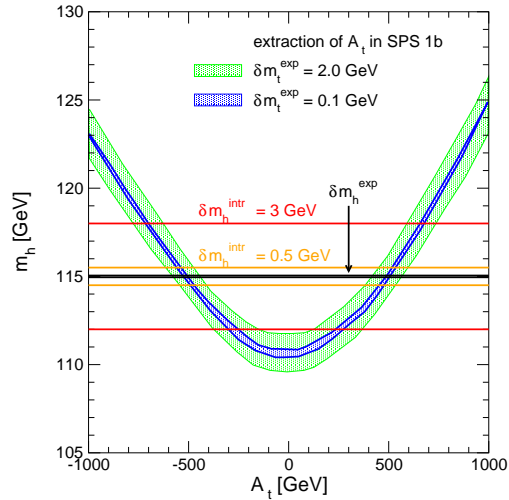}
\caption{Extracting the trilinear coupling $A_t$ from radiative
         corrections to the light MSSM Higgs mass; Ref.~\cite{r16}.}
\label{F5}
\end{center}
\end{figure}
%%%%%%%%%%%%%%%%%
%%%%%%%%%%%%%%%%%
\begin{figure}[h]
\begin{center}
\includegraphics[width=9cm]{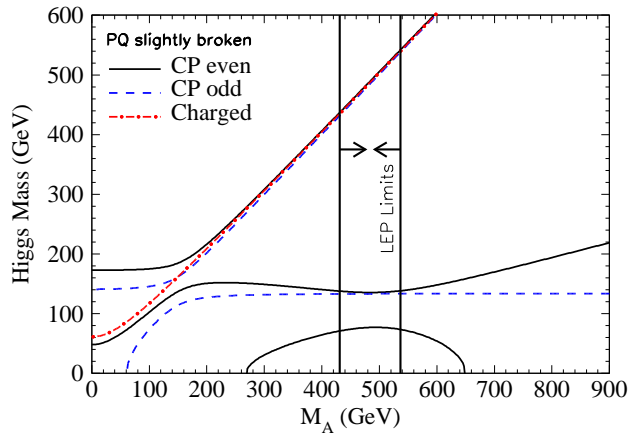}
\caption{Typical Higgs mass spectrum in the non-minimal supersymmetric model NMSSM;
         Ref.~\cite{r17.1}.}
\label{F6}
\end{center}
\end{figure}
%%%%%%%%%%%%%%%%%
the Higgs mass, Fig.~\ref{F5}. For an error on the top mass of $\delta m_t = 100$ MeV,
cf. Ref.~\cite{r16.t},
and an error on the Higgs mass of $\delta m_h = 50$ MeV, cf. Ref.~\cite{r4.2}, $A_t$ 
can be determined at an accuracy of about 10\%.

\vspace{\baselineskip}
\noindent\underline{\it Extended Supersymmetric Theories}  

\vspace{0.5\baselineskip}
A large variety of theories, grand unified theories, string theories,
etc., suggest additional Higgs fields beyond the minimal set in
supersymmetric theories. Adding a complex iso-scalar to the
iso-doublets, an additional pair of scalar and pseudoscalar Higgs
particles is predicted. The axion-type character of the pseudoscalar
boson renders this particle preferentially light. In general, the
standard set of light and heavy Higgs bosons is expected in analogy to
the MSSM, augmented however by a light scalar
and a pseudoscalar Higgs boson, cf. Ref.~\cite{r17.1} and Fig.~\ref{F6}.

If the light pseudoscalar Higgs boson decays to $b$-quarks, a fan of
$b$-jets is expected in Higgs-strahlung as the scalar Higgs bosons may
decay to a pair of light pseudoscalar Higgs bosons, generating at
least four $b$'s in the final state~\cite{r17.2.1}. Nevertheless, a
no-lose theorem for discovering at least one Higgs boson has been
established for ILC while the situation is presently less clear for
LHC~\cite{r17.2.2}.

\subsection{Strong Electroweak Symmetry Breaking} 

Within the Standard Model and its supersymmetric extensions, the Higgs
field is introduced as a fundamental degree of freedom.  Dynamical
electroweak symmetry breaking is rooted in new strong interactions,
not necessarily involving a Higgs boson.  If global symmetries of
these interactions are broken spontaneously, a set of Goldstone bosons
will be generated, such as pions after breaking chiral symmetries in
QCD. By absorbing these Goldstone bosons, longitudinal degrees of
freedom and masses are generated for gauge bosons. Several scenarios
have been developed along this path quite early \cite{r1.4.2,r17.3} as
an alternative to the standard Higgs mechanism and more recently in a
variant responding to the success of the light Higgs picture in
accounting for the high-precision data in the electroweak sector.

%%%%%%%%%%%%%%%%%
\begin{figure}[b]
\begin{center}
\includegraphics[width=8.2cm]{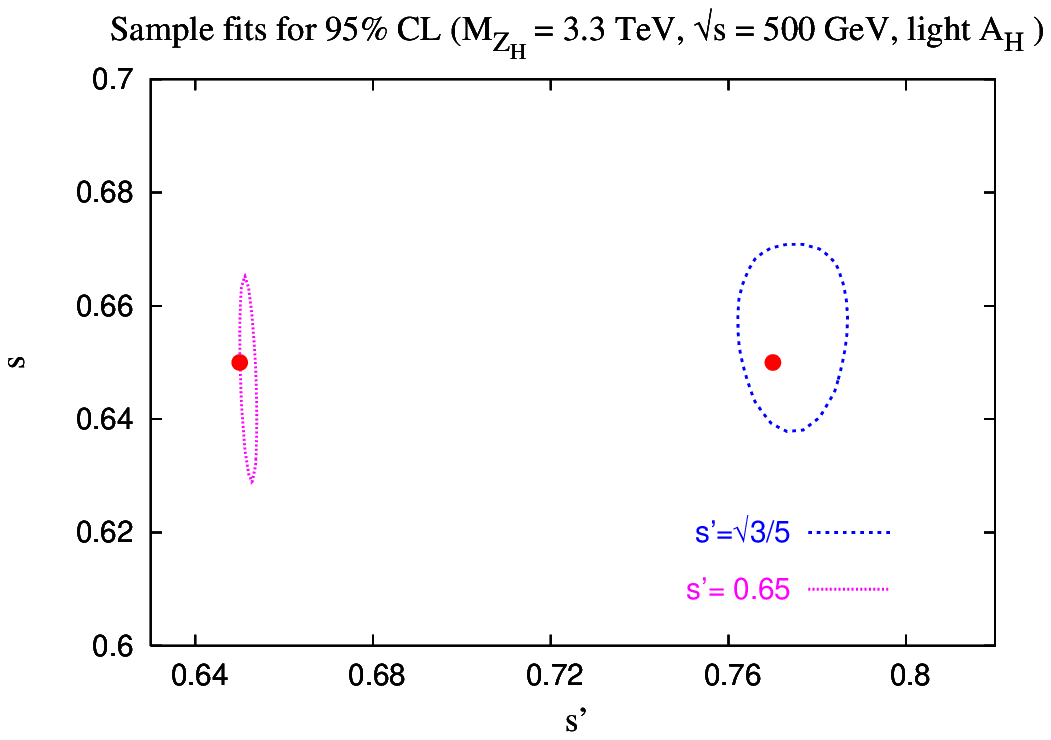}
\includegraphics[width=7.5cm]{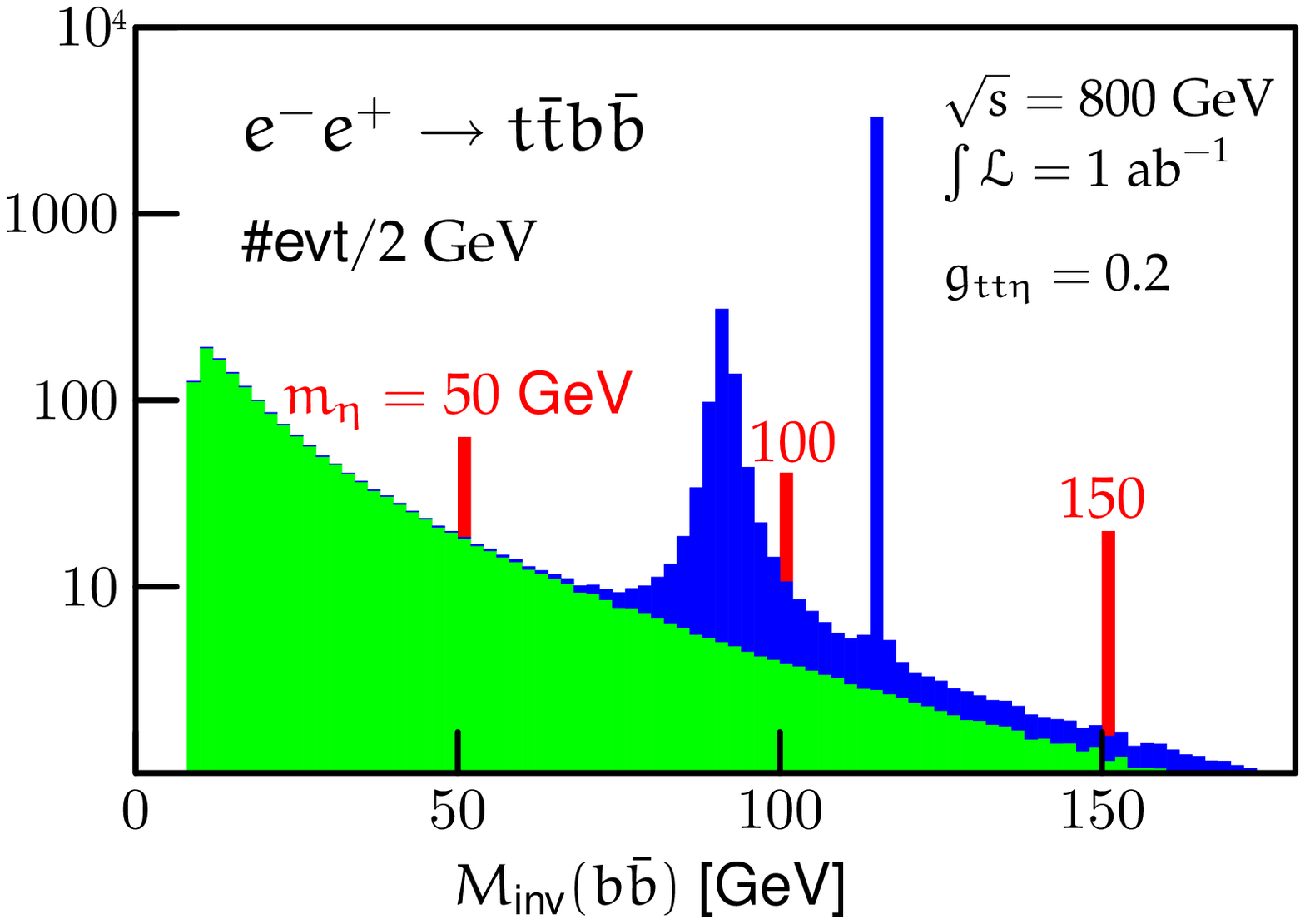}
\caption{Little-Higgs models. Left: Sensitivity of SM processes to LH
         parameters; Ref.~\cite{r18}.  Right: Higgs and pseudoscalar
         boson production; Ref.~\cite{r19}.}
\label{F7}
\end{center}
\end{figure}
%%%%%%%%%%%%%%%%%

\vspace{\baselineskip}
\noindent\underline{\it Little-Higgs Models}    

\vspace{0.5\baselineskip}
These models~\cite{r17.4} are based on new unspecified interactions
that are characterized by a scale $\Lambda$ of order 10 TeV or
more. The breaking of a huge global symmetry, e.g.,
SU(5) $\to$ SO(5), generates a set of pseudo-Goldstone bosons with
properties characterized by the scale $F \sim \Lambda /4\pi$ which is
close to a TeV. Collective breaking of the global symmetry renders the
would-be Goldstone bosons massive; some of the Goldstone bosons however 
acquire mass only at the level of $v\sim F/4\pi$. Interactions in the 
top-quark sector induce a negative mass squared for at least one scalar 
isodoublet, which is thus identified with the standard Higgs degrees, 
while the electroweak mass scale is set by $v=(\sqrt2 G_F)^{-1/2}$.  

While the new (multi-)TeV scalars and vectors may be searched for at
LHC, at ILC their properties can be determined very precisely even if
they remain virtual at the available energies~\cite{r17.5,r18}, cf.\
Fig.~\ref{F7}. Moreover, the entire parameter range of the model,
as expected on general grounds, can be covered in searching for
deviations from the Standard Model predictions in processes such as
$e^+e^- \to f \bar{f}, \ W^+W^-,\  Z H$, and $\gamma\gamma\to H$.  

Little-Higgs models predict a rich spectrum of new particles not only
at the TeV scale, but new states may also be realized at low
scales. Axion-type pseudoscalar bosons may be associated with the
spontaneous breaking of $U(1)$ factors in the extra global
symmetries. These particles have properties analogous to Higgs bosons
\cite{r19}. They are produced parallel to Higgs bosons and their decay
modes may be $b$-jet pairs:
\begin{equation}
e^+e^- \to t \bar{t} \eta \;\; {\rm with}  \;\; \eta \to b \bar{b} \,.
\end{equation}
Thus, instead of one Higgs resonance peak in the invariant $b \bar{b}$
mass in addition to the $Z$ resonance, two peaks would be observed
experimentally, Fig.~\ref{F7}. In $\gamma \gamma$ collisions the two
states could be disentangled by using linearly polarized photon beams;
scalars are generated in collisions of photons with parallel, pseudoscalars 
with perpendicular polarization vectors.

\vspace{\baselineskip}
\noindent\underline{\it Strongly interacting $W,Z$ Bosons}     

\vspace{0.5\baselineskip}
If no Higgs boson will be observed with mass below 1 TeV,
quantum-mechanical unitarity demands strong interactions between the
electroweak gauge bosons, becoming effective at energies $(8\pi/\sqrt2
G_F)^{1/2}\simeq 1.2$
TeV to damp the growth of the amplitudes for (quasi-)elastic $WW$
scattering processes. To achieve compatibility with the $S,T$
parameters extracted from the precision electroweak data at low
energies, the underlying theory must deviate from the template of QCD
as a strongly-interacting theory, which exhibits similar symmetries at
energies below a GeV.  

The new interactions between the electroweak bosons, generically
called $W$, can be expanded in a series of effective interaction terms
with rising dimensions~\cite{r19.1}. Scattering amplitudes are expanded
correspondingly in a series characterized by the energy coefficients
$s/\Lambda_\ast^2$. Demanding CP-invariance and isospin-invariance, as
suggested by the $\rho$ parameter value very close to one, two new
dimension-4 interaction terms must be included in the expansion,
$\mathcal{L}_4 = \alpha_4 \langle W_\mu W_\mu \rangle^2$ and
$\mathcal{L}_5 = \alpha_5 \langle W_\mu W_\nu \rangle ^2$, with
coefficients $\alpha_{4,5} = v^2/\Lambda_{\ast 4,5}^2$ expressed in
the new strong interaction scales $\Lambda_{\ast}$, cf.\
Ref.~\cite{r20}.  To compensate the growth of the scattering
amplitudes in the perturbative expansion, the new contributions must
match the perturbative loop factor $1/16\pi^2$, i.e., the scale
parameters are bounded from above by the value $4\pi v$.  

Quasi-elastic $WW$ scattering,
\begin{equation}
WW \to WW \;\; {\rm and} \;\; ZZ
\end{equation}
can be measured in the processes $e^+e^- \to \bar{\nu} \nu \,WW \;\,
{\rm and} \;\,\bar{\nu} \nu ZZ$. The new interaction terms affect the total cross
sections and the final-state distributions \cite{r20}. The reconstruction 
and separation of $W$ and $Z$ bosons in these analyses is a necessary
condition, which can be fulfilled indeed in the clean environment of a
lepton collider~\cite{r21,r21.1}. Since the impact
%%%%%%%%%%%%%%%%%
\begin{figure}[h]
\begin{center}
\includegraphics[width=8cm]{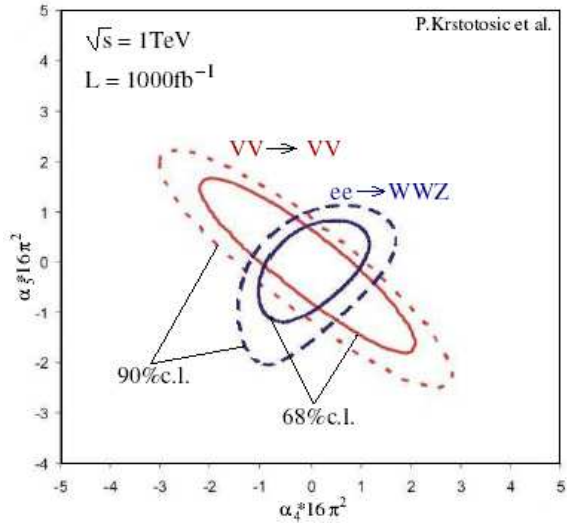}
\caption{Measurement of parameters in new strong interactions of
         electroweak $W$ bosons; Ref.~\cite{r21.1}.}
\label{F8}
\end{center}
\end{figure}
%%%%%%%%%%%%%%%%%
of the new interactions grows with the energy, ILC in the 1 TeV phase
provides the most sensitive instrument for these studies. In fact,
cf.\ Fig.~\ref{F8}, the entire range of $\Lambda_\ast$ values can be
covered experimentally:
\begin{equation}
\Lambda_\ast \leq 4 \pi v \simeq 3 \, {\rm TeV} \, .
\end{equation}                   
The $\Lambda_\ast$ values determine the masses of the resonances
associated with the new interactions. The predictions can be helpful
in the search for these resonances at LHC.

\subsection{Extra Space Dimensions} 

A plethora of different models have been constructed which can break the
electroweak symmetries in scenarios of extra space dimensions. We will
focus on a few characteristic aspects.  

{\it (i)} In Randall-Sundrum models, a scalar {\it radion} field is
introduced to stabilize the distance between the SM and the gravity
brane. Carrying the same quantum numbers as the Higgs field, these two
fields can mix and the properties of the Higgs boson will be
altered \cite{r22.1}. In particular the branching ratio for Higgs decays
to gluon jets may increase dramatically due to dominating radion decays
to gluons, cf. \cite{r22.2}.  

{\it (ii)} {\it Kaluza-Klein states} can affect the $\gamma \gamma$ 
coupling and other loop-induced couplings of the Higgs field. 
Since the $\gamma \gamma$ width of the Higgs particle
can be determined with an accuracy of 2\% in the $\gamma \gamma$
fusion process at a photon-photon collider, the measurement provides
the opportunity to study the particle sector associated with universal
extra dimensions, for example, cf.\ Ref.~\cite{r23}.  

{\it (iii)} Without introducing a scalar Higgs field, electroweak
symmetries can be broken by choosing appropriate {\it boundary
conditions} for the gauge fields in the compactified fifth
dimension. Cancellations which delay unitarity violations at high
energies in $WW$ scattering, are achieved by the exchange of
Kaluza-Klein fields~\cite{r24.1}. Sum rules connect the quartic
couplings of the gauge fields with the couplings between gauge fields
and Kaluza-Klein fields. The Kaluza-Klein states can be searched for
at LHC and ILC~\cite{r24}. At ILC the couplings are expected to be
measured, even for the exchange of virtual Kaluza-Klein fields, quite
accurately.

\section{SUPERSYMMETRY AND UNIFICATION}

If supersymmetry is realized in nature, cf.\ Ref.~\cite{r25}, this
fundamental symmetry will have an impact across all areas in
microscopic physics and cosmology. In the Higgs sector, supersymmetry
would be crucial for generating a light Higgs boson and stabilizing
the electroweak scale in the background of the grand unification and
Planck scales. The contribution of the supersymmetric particle
spectrum to the evolution is essential for the electromagnetic, weak
and strong gauge couplings to approach each other at a high scale, a
necessary condition for the unification of all three gauge
interactions. In addition, local supersymmetry provides a rationale
for gravity by demanding the existence of spin-2 gravitons.  

\begin{figure}[h]
\begin{center}
\includegraphics[width=7cm]{F9a.eps}
\raisebox{5mm}{
\includegraphics[width=7.5cm]{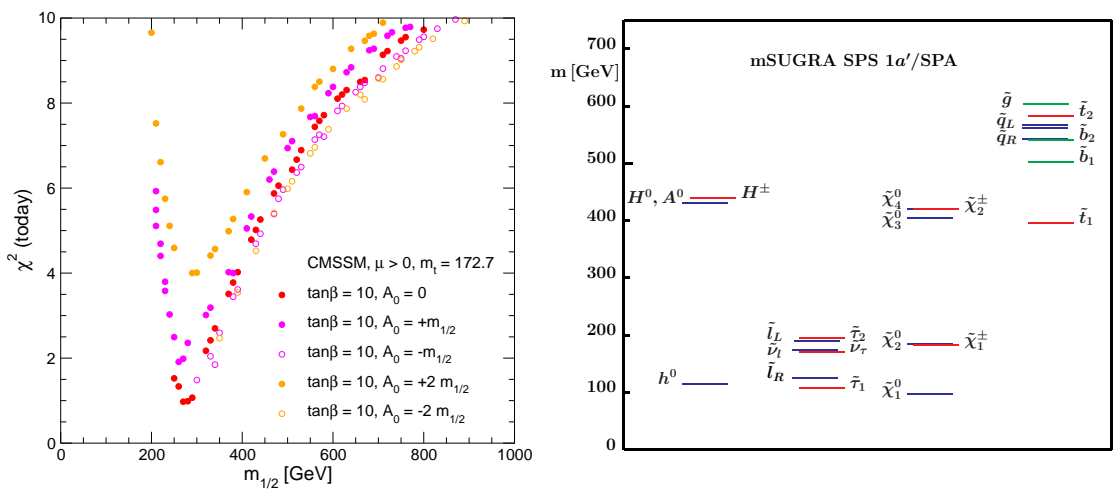}}
\caption{Left: Probability distribution of supersymmetry parameters
         (CMSSM) from precision data for $\tan\beta = 10$;
         Ref.~\cite{r26}. Right: The corresponding mass spectrum near
         the maximum, encoded in the benchmark point SPS1a$'$;
         Ref.~\cite{r27}.}
\label{F9}
\end{center}
\end{figure}

No firm prediction is possible for the mass scale of
supersymmetry. However, for moderate values of the Higgs mixing
parameter $\tan\beta$ a fairly low mass spectrum is indicated in 
the constrained minimal supersymmetric model by combining results 
from radiative 
corrections to electroweak precision observables, $(g_\mu -2)/2$ 
and $b \to s \gamma$, with the measurement of the cold dark matter 
density at WMAP, cf.\ Fig.~\ref{F9}. 
The spectrum corresponding to a parameter set with close to maximal
probability is depicted in Fig.~\ref{F9}. This spectrum had been
chosen as a benchmark set for
%%%%%%%%%%%%%%%%%
%%%%%%%%%%%%%%%%%
a minimal supergravity scenario in the SPS1a$'$ project
\cite{r27}. For the large
$\tan\beta$ range, the quality of the fit is worse and the typical 
mass scale shifts to somewhat larger values. 

The hadron collider LHC and an $e^+e^-$ linear collider are a perfect
tandem for exploring supersymmetry, cf.\ Ref.~\cite{r28}.  The heavy
colored supersymmetric particles, squarks and gluinos, can be
discovered for masses up to 3~TeV with large rates at LHC. Subsequent
cascade decays give access to lower mass particles. The properties of
the potentially lighter non-colored particles, charginos/neutralinos
and sleptons, can be studied very precisely at an $e^+e^-$ linear
collider by exploiting in particular polarization phenomena at such 
a facility. After the properties of the light particles are
determined precisely, the heavier particles can subsequently be
studied in the cascade decays with similar precision.  

Coherent hadron and lepton analyses will provide a comprehensive and
high-precision picture of supersymmetry at the electroweak scale, cf.\
Refs.\cite{r28.1,r29}.  In particular linear-collider experiments will
finally prove supersymmetry to be the origin of the observed new
particles. Moreover, the emerging picture defines, on one side, a
solid basis for the reconstruction of the fundamental supersymmetric
theory near the Planck scale, and for the connection of particle
physics with cosmology on the other side.

\subsection{Properties of Supersymmetric Particles}

For illustration the parameters of the mSUGRA reference point SPS1a$'$
\cite{r27}, a derivative of the Snowmass point SPS1a \cite{r30}, will
be adopted. This point gives a comprehensive picture of the potential
which is offered by coherent analyses at high energy hadron and
$e^+e^-$ colliders. It is characterized by the following values of the
soft parameters at the grand unification scale:
\begin{eqnarray}
  \begin{array}{ll}
    M_{1/2} = 250~{\rm GeV} \qquad & M_0=70~{\rm GeV} \\
    A_0=-300~{\rm GeV} & {\rm sign}(\mu)=+\\
    \tan\beta=10 & \\
  \end{array}
\label{eq:sps1a}
\end{eqnarray}
The universal gaugino mass is denoted by $M_{1/2}$, the scalar mass by
$M_0$ and the trilinear coupling by $A_0$; the sign of the higgsino
mass parameter $\mu$ is chosen positive and $\tan\beta$, the ratio of
the vacuum-expectation values of the two Higgs fields, in the medium
range.  The modulus of the higgsino mass parameter is fixed by
requiring radiative electroweak symmetry breaking so that $\mu
= +396$ GeV.  As shown by the supersymmetric particle spectrum in
Fig.~\ref{F9}, the squarks and gluinos can be studied very well at the
LHC while the non-colored gauginos and sleptons can be analyzed partly
at LHC and in comprehensive and precise form at an $e^+e^-$ linear
collider operating at a total energy up to 1~TeV. 

\vspace{\baselineskip}
\noindent\underline{\it Masses} 

\vspace{0.5\baselineskip}
At LHC, the masses can best be obtained by analyzing edge effects in
the cascade decay spectra, cf.\ Ref.~\cite{r31}. The basic starting
point is the identification of a sequence of two-body decays:
\mbox{$\tilde q_L\rightarrow\tilde\chi^0_2
q\rightarrow\tilde\ell_R\ell q \rightarrow \tilde\chi^0_1\ell\ell q$}.
The kinematic edges and thresholds predicted in the invariant mass
distributions of the two leptons and the jet determine the masses in a
model-independent way.  The four sparticle masses [$\tilde q_L$,
$\tilde\chi^0_2$, $\tilde\ell_R$ and $\tilde\chi^0_1$] are used
subsequently as input for additional decay chains like \mbox{$\tilde
g\rightarrow\tilde b_1 b\rightarrow \tilde\chi^0_2 bb$}, and the
shorter chains \mbox{$\tilde q_R\rightarrow q \tilde\chi^0_1$} and
\mbox{$\tilde\chi^0_4\rightarrow\tilde\ell\ell$}, which all require
the knowledge of the sparticle masses downstream of the
cascades. Residual ambiguities and the strong correlations between the
heavier masses and the LSP mass are resolved by adding the results
from ILC measurements which improve the picture significantly.

\begin{figure}[h]
\begin{center}
\raisebox{2mm}{
\includegraphics[width=8.0cm]{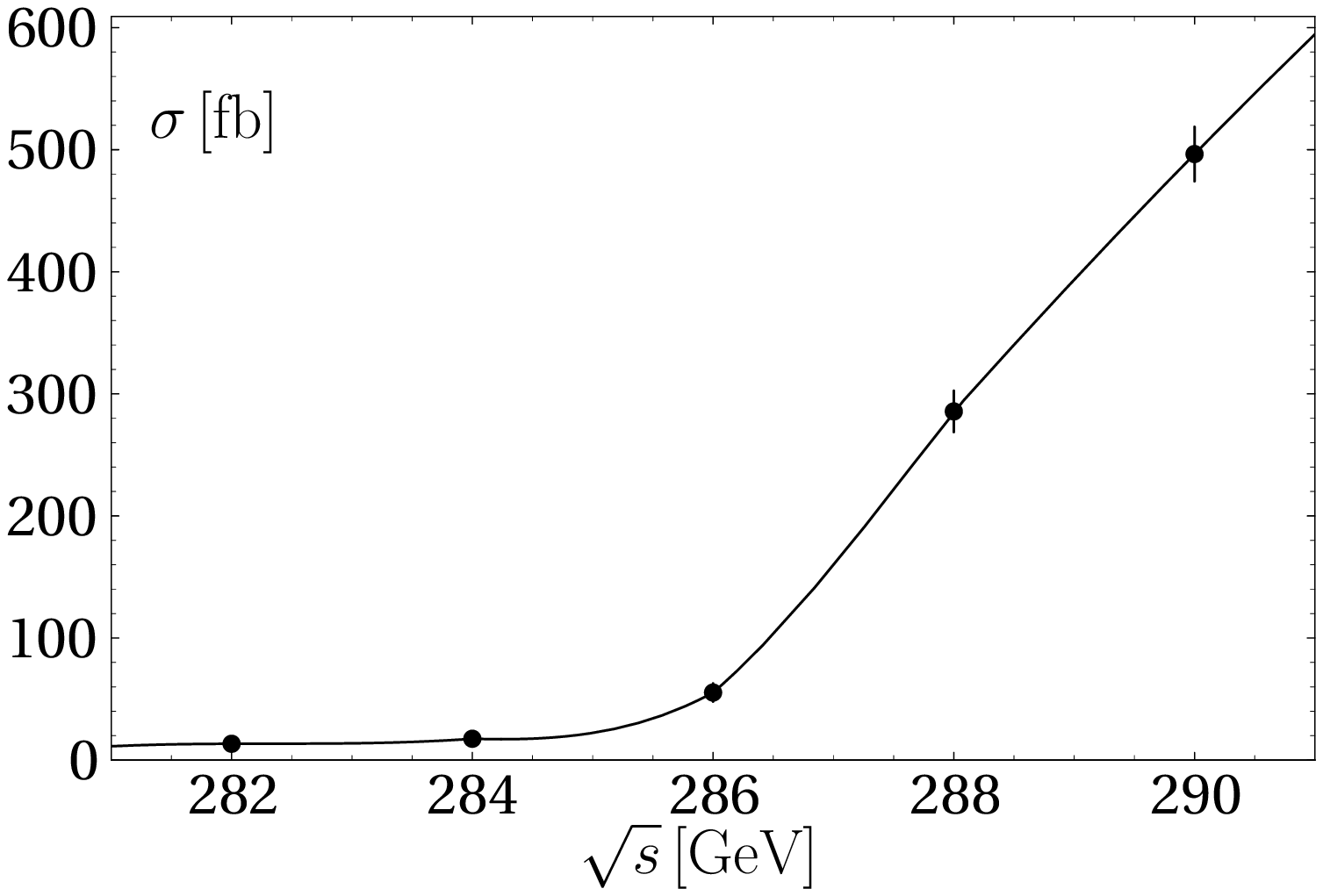}}
\includegraphics[width=7.8cm]{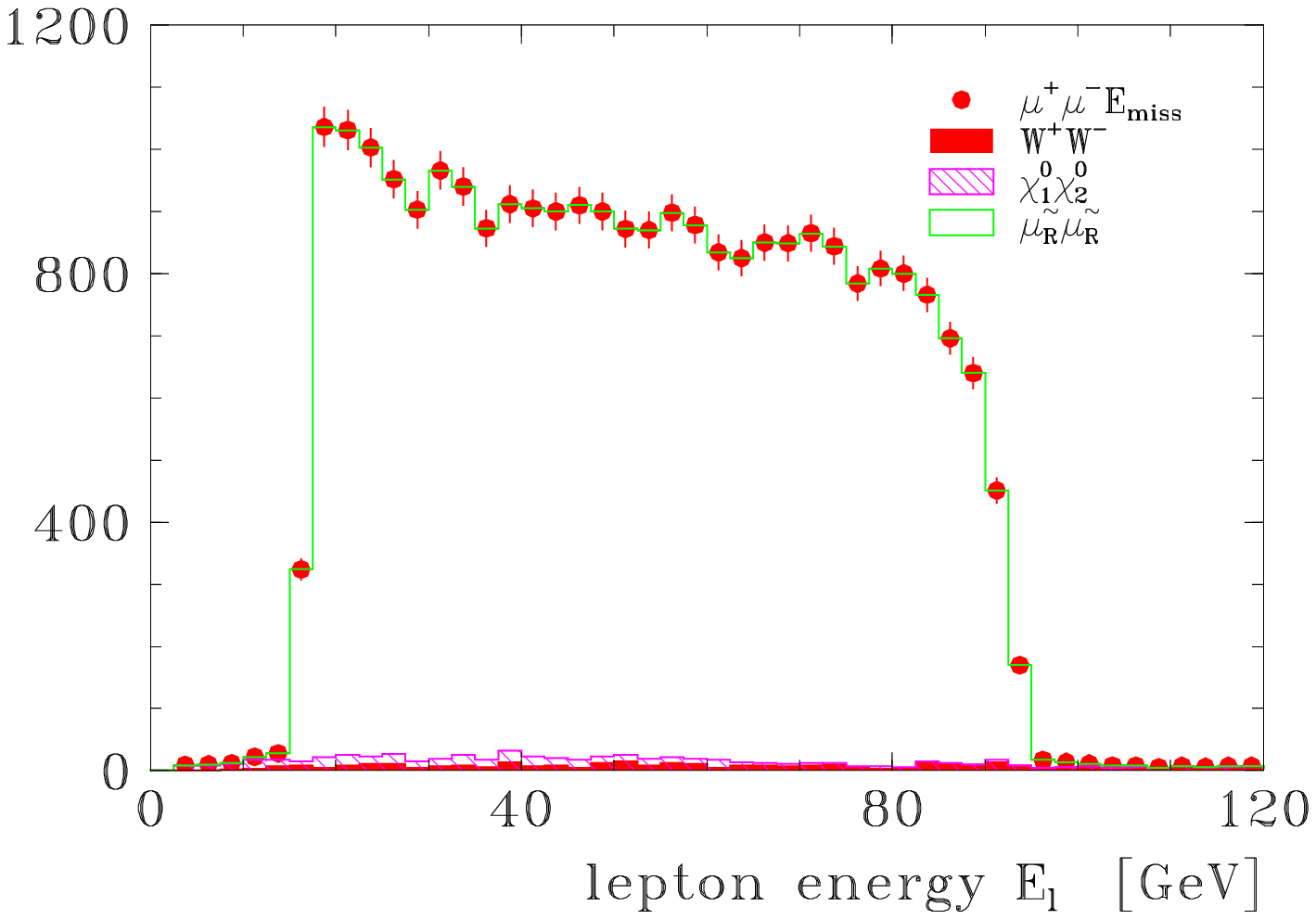}
\caption{Left: Mass measurement in selectron $\tilde{e}^- \tilde{e}^-$ pair production; Ref.~\cite{r32};
         Right: Smuon and neutralino edges in smuon decays; Ref.~\cite{r33}.}
\label{F10}
\end{center}
\end{figure}

At ILC, very precise mass values can be extracted from threshold scans
and decay spectra.  The excitation curves for chargino
$\tilde{\chi}^\pm_{1,2}$ production in S-waves rise steeply with the
velocity of the particles near the thresholds,
\begin{equation}
\sigma \sim \sqrt{s-(\tilde{M}_i+\tilde{M}_j)^2}
\end{equation}
and they are thus very sensitive to their mass values. The same holds
true for mixed-chiral selectron pairs in $e^+e^-\to \tilde e_R^+
\tilde e_L^-$ and for diagonal pairs in $e^-e^-\to \tilde e_R^- \tilde
e_R^-, \; \tilde e_L^- \tilde e_L^-$ collisions, cf.\ Fig.~\ref{F10}.
%%%%%%%%%%%%%%%%%
%%%%%%%%%%%%%%%%%
Other scalar sfermions, as well as neutralinos, are produced generally
in P-waves, with a less steep threshold behavior proportional to the
third power of the velocity. Additional information, in particular on
the lightest neutralino $\tilde{\chi}^0_1$, can be obtained from the
sharp edges of 2-body decay spectra, such as ${\tilde{l}}^-_R \to l^-
{\tilde{\chi}}^0_1$.  Denoting maximum and minimum edge of the decay
lepton spectrum by $E_\pm$, cf.\ Fig.~\ref{F10}, slepton and
${\tilde{\chi}}^0_1$ masses can be derived from
   \begin{eqnarray}
      m_{\tilde\ell} &=& \sqrt{s}\,[E_+E_-]^{\frac{1}{2}}/(E_+ + E_-) 
\nonumber  \\[1mm]
      m_{\tilde{\chi}^0_1} &=& m_{\tilde\ell}\,
                             [1 - 2(E_+ + E_-)/\sqrt{s}]^{\frac{1}{2}}
   \end{eqnarray}
The accuracy in the measurement of the LSP ${\tilde{\chi}}^0_1$ mass
can be improved at ILC by two orders of magnitude compared with LHC.

The values of typical mass parameters and their related measurement
errors are presented in Tab.~\ref{tab:massesA}: ``LHC'' from LHC
analyses and ``ILC'' from ILC analyses; the third column ``LHC+ILC''
presents the corresponding errors if the experimental analyses are
performed coherently, i.e., the light particle spectrum, studied at
ILC with high precision, is used as input set for the LHC analysis.
 \renewcommand{\arraystretch}{1.2}
\begin{table}[h] \footnotesize
\begin{center} $
  \begin{array}{|c|c||c|c||c|}
    \hline
    \ \mbox{Particle} \ &
    \ \ \mbox{Mass}\ \  & \mbox{``LHC''} & \mbox{``ILC''}
                        & \mbox{``LHC+ILC''}\\
    \hline\hline
    h^0                 & 116.9 & 0.25 & 0.05 & 0.05 \\
    H^0                 & 425.0 &      & 1.5  & 1.5  \\
    \hline
    \tilde{\chi}^0_1    &  97.7 & 4.8  & 0.05 & 0.05 \\
    \tilde{\chi}^0_2    & 183.9 & 4.7  & 1.2  & 0.08 \\
    \tilde{\chi}^0_4    & 413.9 & 5.1  & 3-5  & 2.5  \\
    \tilde{\chi}^\pm_1  & 183.7 &      & 0.55 & 0.55 \\ \hline
    \tilde{e}_R         & 125.3 & 4.8  & 0.05 & 0.05 \\
    \tilde{e}_L         & 189.9 & 5.0  & 0.18 & 0.18 \\
    \tilde{\tau}_1      & 107.9 & 5-8  & 0.24 & 0.24 \\ \hline
    \tilde{q}_R         & 547.2 & 7-12 & -    & 5-11 \\
    \tilde{q}_L         & 564.7 & 8.7  & -    & 4.9  \\
    \tilde{t}_1         & 366.5 &      & 1.9  & 1.9  \\
    \tilde{b}_1         & 506.3 & 7.5  & -    & 5.7  \\ \hline
    \tilde{g}           & 607.1 & 8.0  & -    & 6.5  \\ \hline
  \end{array}$ \\
\end{center}
\caption{Accuracies for representative mass measurements
    of SUSY particles in individual LHC, ILC and
    coherent ``LHC+ILC'' analyses
    for the reference point SPS1a$'$ [masses in {\rm GeV}].
    $\tilde q_R$ and $\tilde q_L$ represent the flavors
    $q=u,d,c,s$; cf.\ Ref.~\cite{r28.1}.}
\label{tab:massesA}
\end{table}

\vspace{\baselineskip}\noindent\underline{\it Spins}  

\vspace{0.5\baselineskip}
Determining the spin of new particles is an important measurement
to clarify the nature of the particles and the underlying theory. This
is necessary to discriminate the supersymmetric interpretation of new
particles from other models. A well-known example is the distinction
between supersymmetric theories and theories of universal extra space
dimensions in which new Kaluza-Klein states carry spins different from
supersymmetric particles.  

The measurement of spins in particle cascades at LHC is an
experimental challenge~\cite{r32.1}. Spin measurement at ILC, on the
other hand, is quite easy. The polar angular distribution of smuon
pairs, for example, approaches the characteristic $\sin^2 \theta$ law
for energies sufficiently above threshold. The smuons can be
reconstructed up to a discrete ambiguity; false solutions in the
reconstruction generate a flat background underneath the
signal \cite{r1.1}.  

\vspace{\baselineskip}
\noindent\underline{\it Mixings}   

\vspace{0.5\baselineskip}
Mixing parameters must be extracted from measurements of cross
sections and polarization asymmetries. In the production of charginos
and neutralinos, both diagonal or mixed pairs can be exploited:
$e^+e^- \rightarrow {\tilde{\chi}^+_i}{\tilde{\chi}^-_j}$
\mbox{[$i$,$j$ = 1,2]} and ${\tilde{\chi}^0_i} {\tilde{\chi}^0_j}$
[$i$,$j$ = 1,$\dots$,4].  The production cross sections for charginos
are binomials in $\cos\,2\phi_{L,R}$, the mixing angles rotating
current to mass eigenstates. Using polarized electron and positron
beams, the mixings can be determined in a model-independent way
\cite{r34.1,r34.2}.  

The same methods can be applied to determine the mixings in the scalar
sfermion sector. The production cross sections for stop particle
pairs, $e^+e^- \to \tilde{t}_i \tilde{t}^c_j$ \mbox{[$i$,$j$ =
1,2]}, depend on the mixing parameters
$\cos\!/\!\sin{2\theta_{\tilde{t}}}$ which can be determined with high
accuracy by making use of polarized electron beams \cite{r35}.  

The measurement of the discrete quantum numbers of scalar sfermions is
another basic process. Using polarized electron and positron beams, the
L/R quantum numbers of selectrons and positrons can be identified
unambiguously even if the masses are nearly degenerate \cite{r2}.

\vspace{\baselineskip}
\noindent\underline{\it Couplings}  

\vspace{0.5\baselineskip}
Supersymmetry predicts the identity of Yukawa and gauge couplings
among particle partners, in generic notation,
\begin{equation}
V \tilde{V} \tilde{V} = VVV \;\; {\rm and} \;\;  F \tilde{F} \tilde{V} = FFV 
\end{equation}
for gauge bosons $V$ and gauginos $\tilde{V}$, and for fermions $F$
and their scalar partners $\tilde{F}$. These fundamental relations can
be studied experimentally in pair production of charginos and
neutralinos which is partly mediated by the exchange of sneutrinos and
selectrons in the $t$-channel, as well as selectron and sneutrino pair
production which is partly mediated by neutralino and chargino
$t$-channel exchanges.  

\begin{figure}[h]
\begin{center}
\hspace*{-10mm}
\includegraphics[width=7cm]{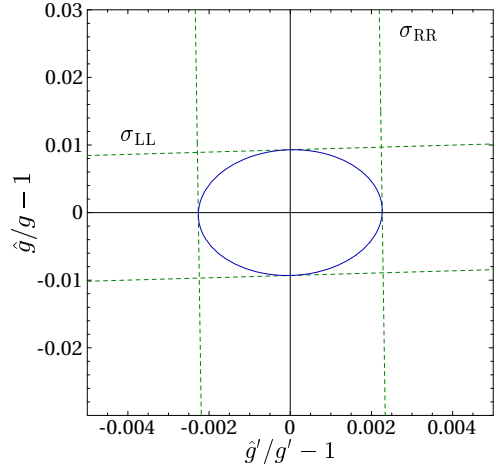}
\caption{Relating the SU(2) and U(1) $\ell \tilde{\ell} \tilde{V}$ Yukawa couplings 
         experimentally to the corresponding gauge couplings $\ell \ell V$ 
         in selectron pair production; Ref.~\cite{r32}.}
\label{F11}
\end{center}
\end{figure}

An example is presented in Fig.~\ref{F11} for the sensitivity which
can be achieved at ILC in testing the identity of Yukawa and gauge
couplings in selectron pair production.
%%%%%%%%%%%%%%%%%
%%%%%%%%%%%%%%%%%
The separation of the electroweak SU(2) and U(1) couplings is possible
if polarized electron beams are available. At the end of course an
overall analysis is required which takes into account the measurements
of the masses and gaugino/higgsino mixing parameters of the neutralinos 
exchanged in the $t$-channel.

\subsection{Fundamental Supersymmetric Theory}

The measurements described in the previous section provide the initial
values for the evolution of the gauge couplings and the soft SUSY
breaking parameters in the Lagrangian to the grand unification scale,
cf.\ Ref.~\cite{r36}, where in many scenarios the fundamental
supersymmetric theory is defined. The values at the electroweak scale
are connected to the fundamental parameters at the GUT scale $M_U$ by
the renormalization group equations; to leading order, 
\begin{center}
\noindent
\begin{tabular}{ll}
 gauge couplings &: $\alpha_i = Z_i \, \alpha_U$      \\
 gaugino masses  &: $M_i = Z_i \, M_{1/2}$            \\
 scalar masses   &: $M^2_{\tilde{j}} = M^2_0 + c_j M^2_{1/2} +
                     \sum_{\beta=1}^2 c'_{j \beta} \Delta M^2_\beta$  \\
 trilinear  couplings &:  $A_k = d_k A_0   + d'_k M_{1/2}$            \\     
\end{tabular}
\end{center}
The index $i$ runs over the gauge groups $i=$ SU(3), SU(2), U(1).
To this order, the gauge couplings, and the gaugino and scalar mass
parameters of soft supersymmetry breaking depend on the $Z$
transporters $Z_i^{-1} = 1 + b_i \alpha_U / (4 \pi)
\log(M_U^2/M_Z^2)$. The scalar mass parameters $M^2_{\tilde{j}}$
depend also on the Yukawa couplings.  Beyond these approximate
solutions, the evolution equations have been solved numerically.  

\vspace{\baselineskip}
\noindent\underline{\it Gauge Coupling Unification}  

\vspace{0.5\baselineskip}
Measurements of the gauge couplings at the electroweak scale support
very strongly the unification of the couplings \cite{r36.g}
at a scale $M_U \simeq
2\times 10^{16}$~GeV, with a precision at the per-cent level.
%%%%%%%%%%%%%%%%%
\begin{figure}[h]
\begin{center}
\hspace*{-10mm}
\unitlength 1mm
\begin{picture}(0,50)
\put(-70,-65){\includegraphics[width=10cm]{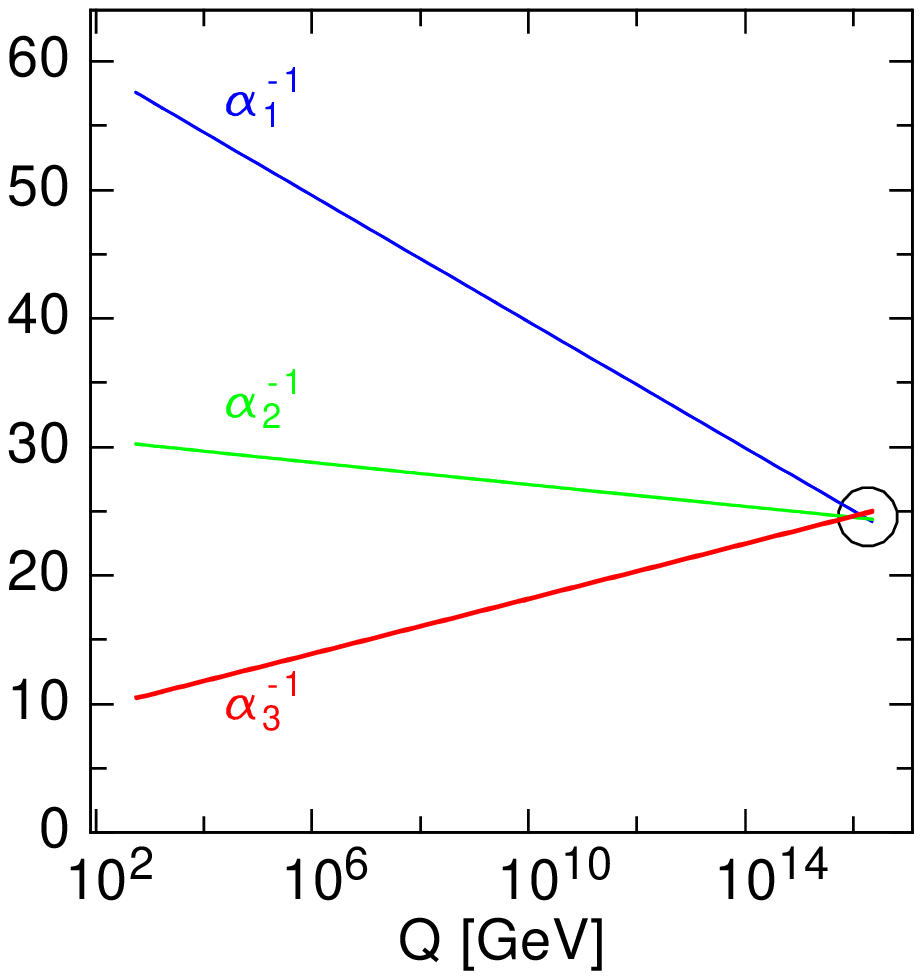}}
\put(0,-65){\includegraphics[width=10cm]{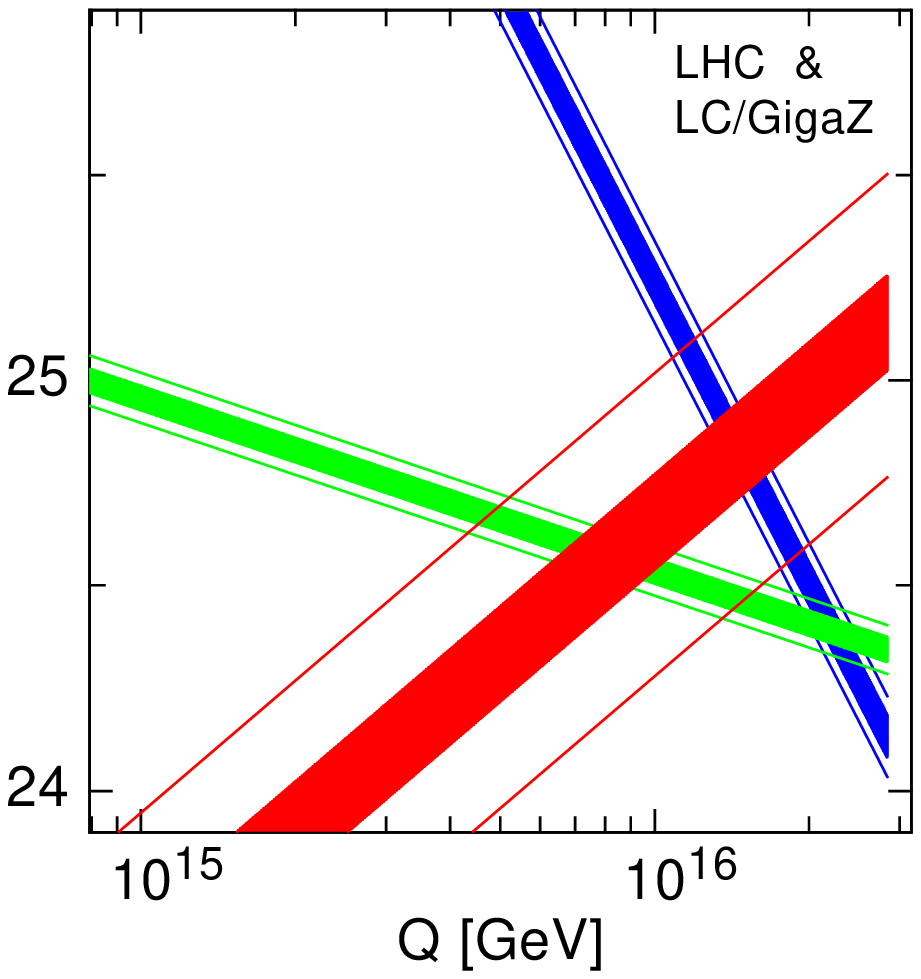}}
\end{picture}
\caption{Gauge coupling unification at GigaZ; Ref.~\cite{r28.1}.}
\label{F12}
\end{center}
\end{figure}
%%%%%%%%%%%%%%%%%
The couplings do not meet exactly, cf.\ Fig.~\ref{F12} and
Tab.~\ref{tab:gauge}, most evident after taking into account results
from GigaZ runs.  The differences are to be attributed to
high-threshold effects at the unification scale $M_U$, and the
quantitative evaluation implies important constraints on the particle
content at $M_U$. 

\begin{table*}[tb]
\centering
\begin{tabular}{|c||c|c|}
\hline
 & Present/``LHC'' & GigaZ/``LHC+ILC'' \\
\hline \hline 
$M_U$ & $(2.36 \pm 0.06)\cdot 10^{16} \, \rm {GeV}$ &
           $ (2.360 \pm 0.016) \cdot 10^{16} \, \rm {GeV}$ \\[0.5mm]
$\alpha_U^{-1}$ & $  24.19 \pm 0.10 $ &  $ 24.19 \pm 0.05 $\\[0.5mm] \hline 
$\alpha_3^{-1} - \alpha_U^{-1}$ & $0.97 \pm 0.45$ & $0.95 \pm 0.12$ \\[0.5mm]
\hline
\end{tabular}
\caption{Precision in extracting the unified gauge coupling
         $\alpha_U$, derived from the meeting point of $\alpha_1$ with
         $\alpha_2$, and the strong coupling $\alpha_3$ at the GUT
         scale $M_U$. The columns demonstrate the results for the
         expected precision from LEP and LHC data, as well as the
         improvement due to a GigaZ linear collider analysis, cf.\
         Ref.~\cite{r28.1}.}
\label{tab:gauge}
\end{table*}

\vspace{\baselineskip}
\noindent\underline{\it Gaugino and Scalar Mass Parameters}  

\begin{figure}[h]
\begin{center}
%% \hspace*{-10mm}
%% \unitlength 1mm
%% \begin{picture}(0,58)
%% \put(-65,-65){\includegraphics[width=13cm]{F13a.eps}}
%% \put(10,-65){\includegraphics[width=13cm]{F13b.eps}}
%% \end{picture}
\setlength{\unitlength}{1mm}
\begin{picture}(150,75)(5,10)
  \put( 0,-70){\includegraphics[width=165mm]{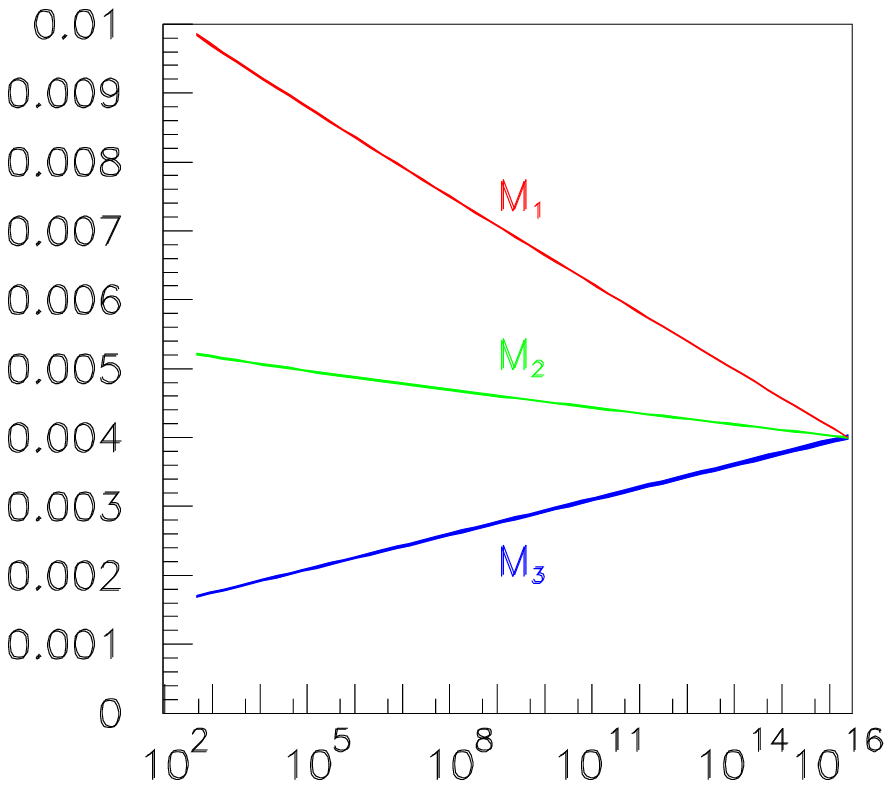}}
  \put(85,-70){\includegraphics[width=165mm]{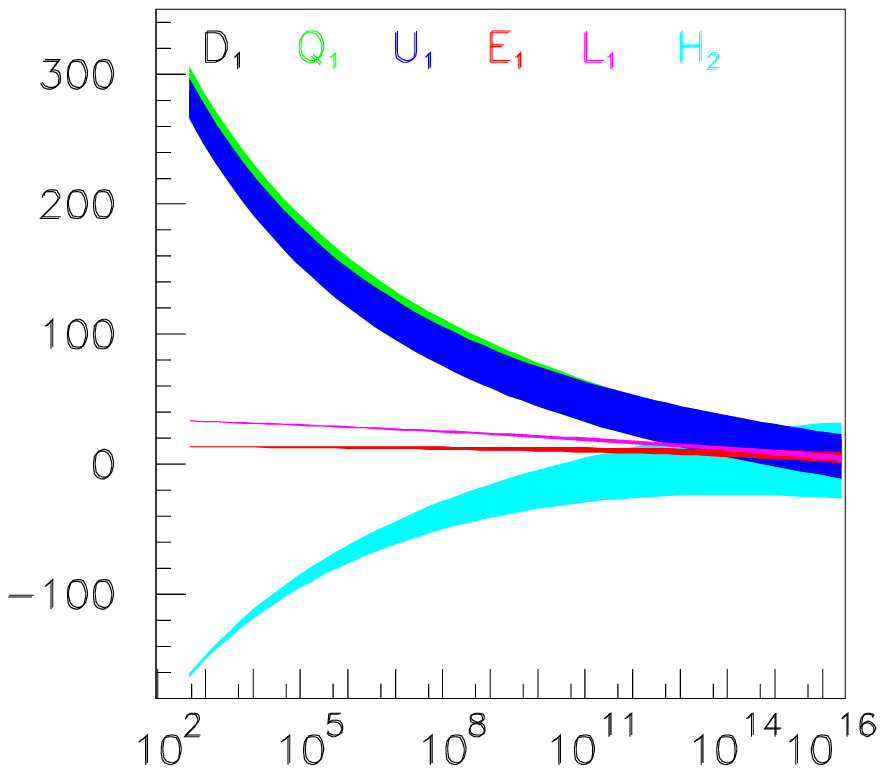}}
  \put(44, 30){\includegraphics[height=7mm,width=8mm]{box.eps}}
  \put(44, 48){\includegraphics[height=7mm,width=8mm]{box.eps}}
  \put(44, 62){\includegraphics[height=7mm,width=8mm]{box.eps}}
  \put(-5,53){\begin{sideways}\large $1/M_i~[{\rm GeV}^{-1}]$ \end{sideways}}
  \put(85,52){\begin{sideways}\large $M^2_{\tilde f}~[10^3\,{\rm GeV}^2]$ \end{sideways}}
  \put( 60,10){\large $Q~[{\rm GeV}]$}
  \put(145,10){\large $Q~[{\rm GeV}]$}
  \put(44, 32){\color{blue}\large $M_3^{-1}$ }
  \put(44, 49){\color{green}\large $M_2^{-1}$ }
  \put(44, 63){\color{red}\large $M_1^{-1}$ }
\end{picture}
\caption{Unification of gaugino and scalar mass parameters; Ref.~\cite{r28.1}.}
\label{F13}
\end{center}
\end{figure}

\vspace{0.5\baselineskip}
The results for the evolution of the mass parameters from the electroweak
scale to the GUT scale $M_U$ are shown in Fig.~\ref{F13}.
On the left of Fig.~\ref{F13} the evolution is presented for the
gaugino parameters $M^{-1}_i$. 
%%%%%%%%%%%%%%%%%
%%%%%%%%%%%%%%%%%
The model-independent reconstruction of the fundamental parameters and
the test of universality in the SU(3)$\times$SU(2)$\times$U(1) group
space are clearly under excellent control.  

In the same way the evolution of the scalar mass parameters can be
studied, presented on the right of Fig.~\ref{F13} for the first/second
generation.  While the slepton parameters can be determined very
precisely, the accuracy deteriorates for the squark parameters and
the Higgs parameter $M^2_{H_2}$.  

The evolution of the scalar mass parameters is quite distinct from scenarios
in which supersymmetry is broken by a different mechanism. A typical example
is gauge mediated supersymmetry breaking GMSB where regularities are predicted
at an intermediate energy scale but extrapolations to Planck scale energies
lead to markedly non-universal mass parameters \cite{r36}. Thus the microscopic 
picture of supersymmetry breaking can be explored this way experimentally.     

These examples demonstrate that high-precision experiments at
high-energy colliders allow us to reconstruct physical scenarios near
the Planck scale.  They shed light on a domain where in many
theoretical approaches the roots of physics are located including
gravity.

\subsection{Left-Right Symmetric Extension}

The complex structure observed in the neutrino sector requires the
extension of the minimal supersymmetric Standard Model MSSM, e.g., by
a superfield including the right-handed neutrino field and its scalar
partner.  If the small neutrino masses are generated by the seesaw
mechanism \cite{r38.1}, a similar type of spectrum is induced in the scalar
sneutrino sector, splitting into light TeV-scale and very heavy
masses.  The intermediate seesaw scales will affect the evolution of
the soft mass terms which break the supersymmetry at the high (GUT)
scale, particularly in the third generation with large Yukawa
couplings.  This will provide us with the opportunity to measure,
indirectly, the intermediate seesaw scale of the third generation
\cite{r38.2}. 

If sneutrinos are lighter than charginos and the second lightest
neutralino, as encoded in SPS1a$'$, they decay only to invisible $\nu
\tilde{\chi}_1^0$ final states.  However, in this configuration
sneutrino masses can be measured in chargino decays to sneutrinos and
leptons.  These two-particle decays develop sharp edges at the
endpoints of the lepton energy spectrum for charginos produced in
$e^+e^-$ annihilation.  Sneutrinos of all three generations can be
explored this way \cite{r38.2}.  The errors for the first and second
generation sneutrinos are expected at the level of 400 MeV, doubling
for the more involved analysis of the third generation.  

The measurement of the seesaw scale can be illustrated in an SO(10)
model~\cite{r37} in which the Yukawa couplings in the neutrino sector
are proportional to the up-type quark mass matrix. The masses of the
physical right-handed Majorana neutrinos are hierarchical, very
roughly $\propto m^2_{\rm up}$, and the mass of the heaviest neutrino
is given by $M_{\RR_3} \sim m^2_t / m_{\nu_3}$ which, for $m_{\nu_3}
\sim 5 \times 10^{-2}$ eV, amounts to $\sim 6 \times 10^{14}$ GeV,
i.e., a value close to the grand unification scale $M_U$.

\begin{figure}[h]
\begin{center}
\hspace*{-10mm}
\includegraphics[width=5.5cm]{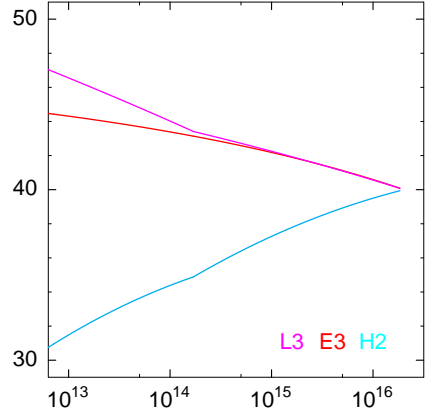}
\caption{Impact of the heavy right-handed neutrino mass on the
         evolution of the scalar mass parameters in LR symmetric
         theories; Refs.~\cite{r36,r38.2}.}
\label{F14}
\end{center}
\end{figure}

Since the $\nu_{\rm R}$ is unfrozen only beyond $Q=M_{\nu_{\rm R}}$ the 
impact of the LR extension becomes visible in the evolution of the scalar 
mass parameters only at very high scales.  The effect of $\nu_{\rm R}$ can 
be manifest only in the third generation where the Yukawa coupling is
large enough; the evolution in the first two generations can thus be
used to calibrate the assumption of universality for the scalar mass
parameters at the unification scale \cite{r36}.  In Fig.~\ref{F14} the
evolution of the scalar mass parameters in the third generation and
the Higgs mass parameter are displayed. The lines
%%%%%%%%%%%%%%%%%
%%%%%%%%%%%%%%%%%
include the effects of the right-handed neutrino which induce the
kinks. Only the picture including $\nu_{\rm R}$,
$\tilde{\nu}_{\rm R}$ is compatible with the assumption of
unification.

The kinks in the evolution of $M^2_{{\tilde L}_3}$ shift the physical
masses [squared] of the ${\tilde{\tau}_{\rm L}}$ and ${\tilde{\nu}_{\tau
{\rm L}}}$ particles of the third generation by the amount
$\Delta_{\nu}[M_\RR]$ compared with the slepton masses of the first
two generations. The measurement of
\begin{equation} 
\Delta_{\nu}[M_{{\rm R}_3}] \propto m_{\nu_3} M_{\RR_3}
\log(M^2_{\rm GUT}/M^2_{\RR_3}) 
\end{equation}
can be exploited to determine the neutrino seesaw scale of the third
generation~\cite{r38.2},
\begin{equation}
 M_{\RR_3} = 3.7 \; {\rm{to}} \; 6.9 \times 10^{14} {\rm GeV}
\end{equation}
in the LR extended SPS1a$'$ scenario with an initial value of $6
\times 10^{14}$ GeV. Thus, this analysis provides us with a unique
estimate of the high-scale $\nu_{\rm R}$ seesaw mass parameter
$M_{\RR_3}$.

\subsection{Split Supersymmetry}

For a successful unification of forces at the GUT scale the sfermion
mass scale $M_0$ is irrelevant, since each generation of
sfermions furnishes a complete representation of SU(5) [or SO(10), if
right-handed sneutrinos are included].  Likewise, the dark-matter
prediction of the MSSM and its extensions does not rely on the value
of $M_0$, but rather on the existence of a conserved discrete
quantum number, R~parity.  These facts are compatible with the speculation
that the sfermion mass scale may actually be much higher than the
gaugino mass scale, effectively removing all scalar partners of the
matter fields and the extra heavy Higgs states of the MSSM from the
low-energy spectrum~\cite{r37.1}.  

Such a scenario has the attractive feature that sources of flavor
violation besides CKM mixing are naturally absent, removing the
requirement of sfermion-mass degeneracy from the mechanism of
supersymmetry breaking.  On the other hand, the Higgs potential is
fine-tuned as in the non-supersymmetric SM.  Explanations for this
fact may be traced back to the property of string theories to provide
a huge landscape of acceptable vacuum states.  The quantum-mechanical
stability of the Higgs-field ground state would be related to the
quantum-mechanical stability of the vacuum energy, i.e., the
cosmological constant.

With a sufficiently high sfermion mass scale, e.g., $M_0 \sim
10^9\;\textrm{GeV}$, the gluino acquires a macroscopic lifetime and,
for the purpose of collider experiments, behaves like a massive,
stable color-octet parton.  This leads to characteristic signatures at
LHC.  Detection of such a particle is possible up to $m_{\tilde g} =
1-2\;\textrm{TeV}$~\cite{r37.2,r37.3}.  The Higgs boson mass is
expected to be above the conventional MSSM mass range, so four-fermion
decays of the Higgs particle dominate over two-fermion decays.  However, 
due to the absence of cascade decays, production of the non-colored 
gauginos and higgsinos at LHC proceeds only via electroweak annihilation
processes, and the production rates are thus considerably suppressed
compared to conventional MSSM scenarios.  

In this situation, the analysis of chargino and neutralino
pair-production at ILC provides the information necessary to deduce
the supersymmetric nature of the model.  Extracting the values of
chargino/neutralino Yukawa couplings, responsible for the mixing of
gaugino and higgsino states, reveals the anomalous effects due to the
splitting of gaugino and sfermion mass scales~\cite{r37.2}.
Furthermore, these parameters determine the higgsino content of the
LSP and thus the relic dark-matter density predicted by the Split
Supersymmetry model~\cite{r37.5}.

\subsection{String Effective Theories}

Heterotic string theories give rise to a set of 4-dimensional dilaton
$S$ and moduli $T$ superfields after compactification.  The vacuum
expectation values of $S$ and $T$, generated by genuinely
non--perturbative effects, determine the soft supersymmetry breaking
parameters.  

The properties of the supersymmetric theories are quite different for
dilaton and moduli dominated scenarios, quantified by the mixing angle
$\theta$. This angle $\theta$ characterizes the $\tilde S$ and $\tilde
T$ components of the wave function of the Goldstino, which is
associated with the breaking of supersymmetry.  The mass scale is set
by the second parameter of the theory, the gravitino mass $m_{3/2}$.
%%%%%%%%%%%%%%%%%
\begin{figure}[h]
\begin{center}
\includegraphics[width=7cm,angle=270]{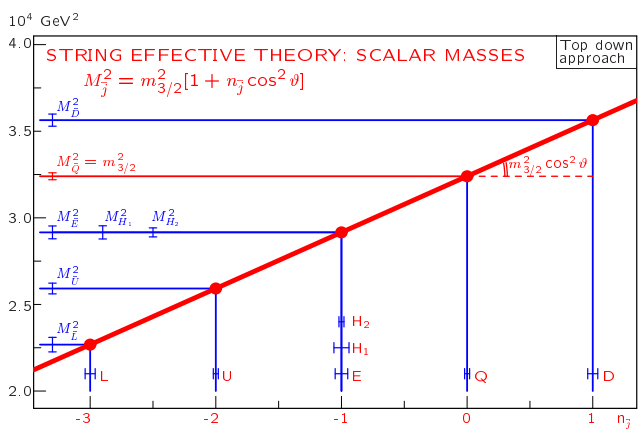}
\vspace{-10mm}
\caption{The linear relation between integer modular weights and scalar mass parameters
         in string effective theories; cf.\ Ref.~\cite{r36}.}
\label{F15}
\end{center}
\end{figure}
%%%%%%%%%%%%%%%%%

In leading order, the masses \cite{r37.0} are given by
\begin{eqnarray}
M_i &=&  - g_i^2 m_{3/2} \langle S \rangle {\sqrt{3} \sin \theta} + ...  \\
M_{\tilde{j}}^2 &=& m^2_{3/2} \left(  1 + n_j \cos^2 \theta \right) + ...
\end{eqnarray}
for the gaugino sector and the scalar sector, respectively.  A dilaton
dominated scenario, $\sin \theta \to 1$, leads to universal boundary
conditions of the soft supersymmetry breaking parameters.  On the
other hand, in moduli dominated scenarios, $\cos\theta \to 1$, the
gaugino mass parameters are universal, but universality is not
realized for the scalar mass parameters. The breaking is characterized
by integer modular weights $n_j$ which quantify the couplings between
the matter and the moduli fields. Within one generation significant
differences between left and right field components and between
sleptons and squarks can occur.  

The results \cite{r36} for the analysis of a mixed dilaton/moduli
superstring scenario with dominating dilaton component, $\sin^2 \theta
= 0.9$, and with different couplings of the moduli field to the (L,R)
sleptons, the (L,R) squarks and to the Higgs fields corresponding to
the O--I representation $n_{L_i} = -3$, $n_{E_i} = -1$, $n_{H_1}
=n_{H_2}=-1$, $n_{Q_i} = 0$, $n_{D_i} = 1$ and $n_{U_i} = -2$, are
presented in Fig.~\ref{F15}.  The gravitino mass is chosen to be
180~GeV in this analysis.
Given this set of superstring induced parameters, the evolution of the
gaugino and scalar mass parameters can be exploited to determine the
modular weights $n$. Fig.~\ref{F15} demonstrates
how stringently this theory can be tested by analyzing
the integer character of the entire set of modular weights. 

Thus, high-precision measurements at high energy proton and lepton
colliders may provide access to crucial derivative parameters
in string theories.

\subsection{Intermediate Gauge Bosons}

Gauge bosons at the intermediate TeV scale are motivated by many
theoretical approaches, cf.\ Ref.~\cite{r37.6}.  The breaking of GUT
theories, based on SO(10) or E(6) symmetries for example, may leave
one or several U(1) remnants unbroken down to TeV energies, before the
symmetry is reduced finally to the SM symmetry:
\begin{eqnarray}
{\rm{SO(10)}} &\to& {\rm{SM}}\times{\rm{U(1)}}                           \\
{\rm{E(6)}}   &\to& {\rm{SO(10)}}\times{\rm{U(1)}} \to 
                  {\rm{SM}}\times{\rm{U(1)}}\times{\rm{U(1)}} \nonumber  \\
              &\to& {\rm{SM}}\times{\rm{U(1)}}   
\end{eqnarray}
The final U(1) remnant of E(6) is a linear combination $\chi$, $\psi$
or $\eta$ of the U(1)'s generated in the two-step symmetry breaking
mechanism.  

%%%%%%%%%%%%%%%%%
\begin{figure}[h]
\begin{center}
\includegraphics[height=7.8cm]{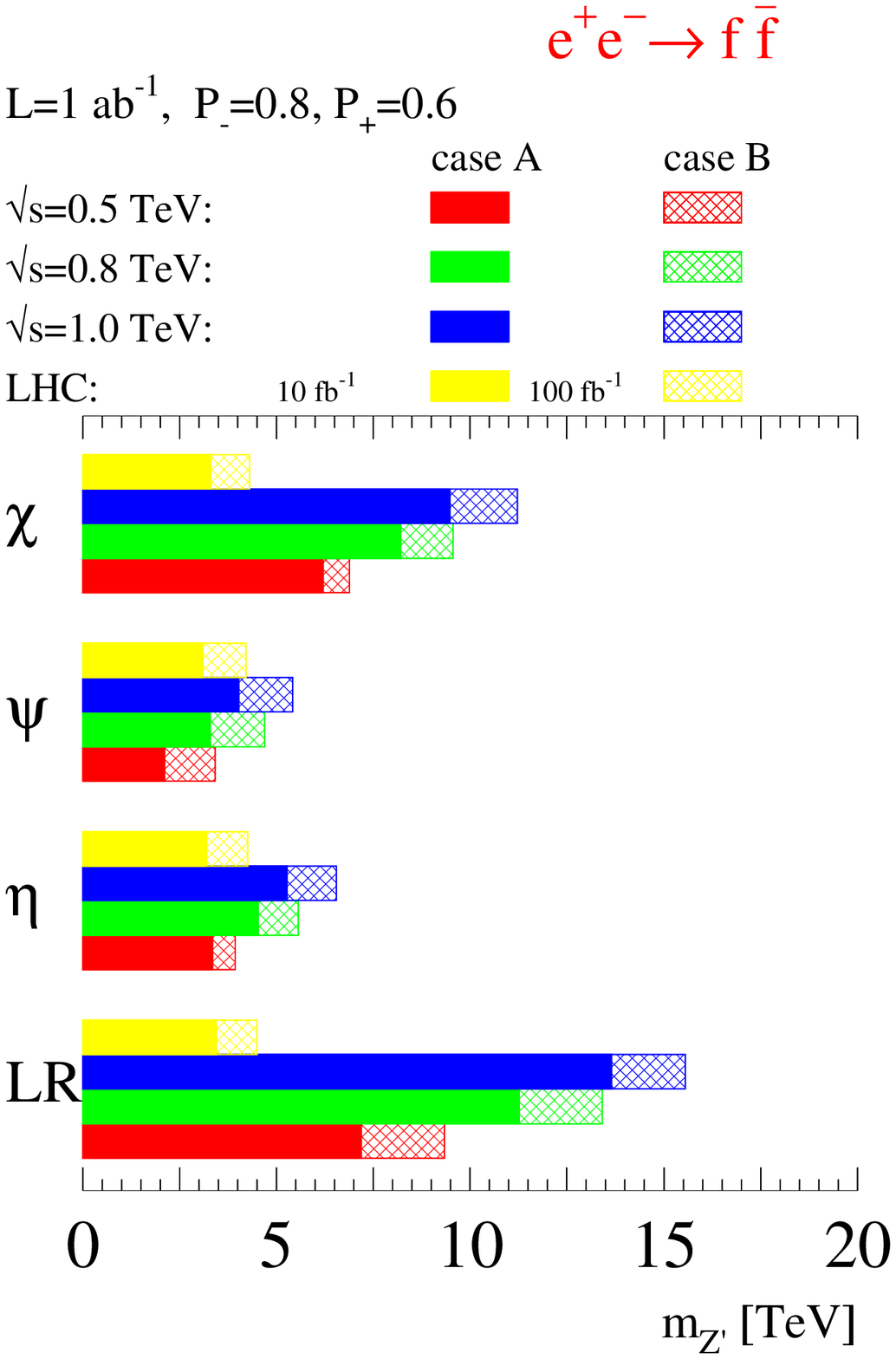}
\includegraphics[height=7cm]{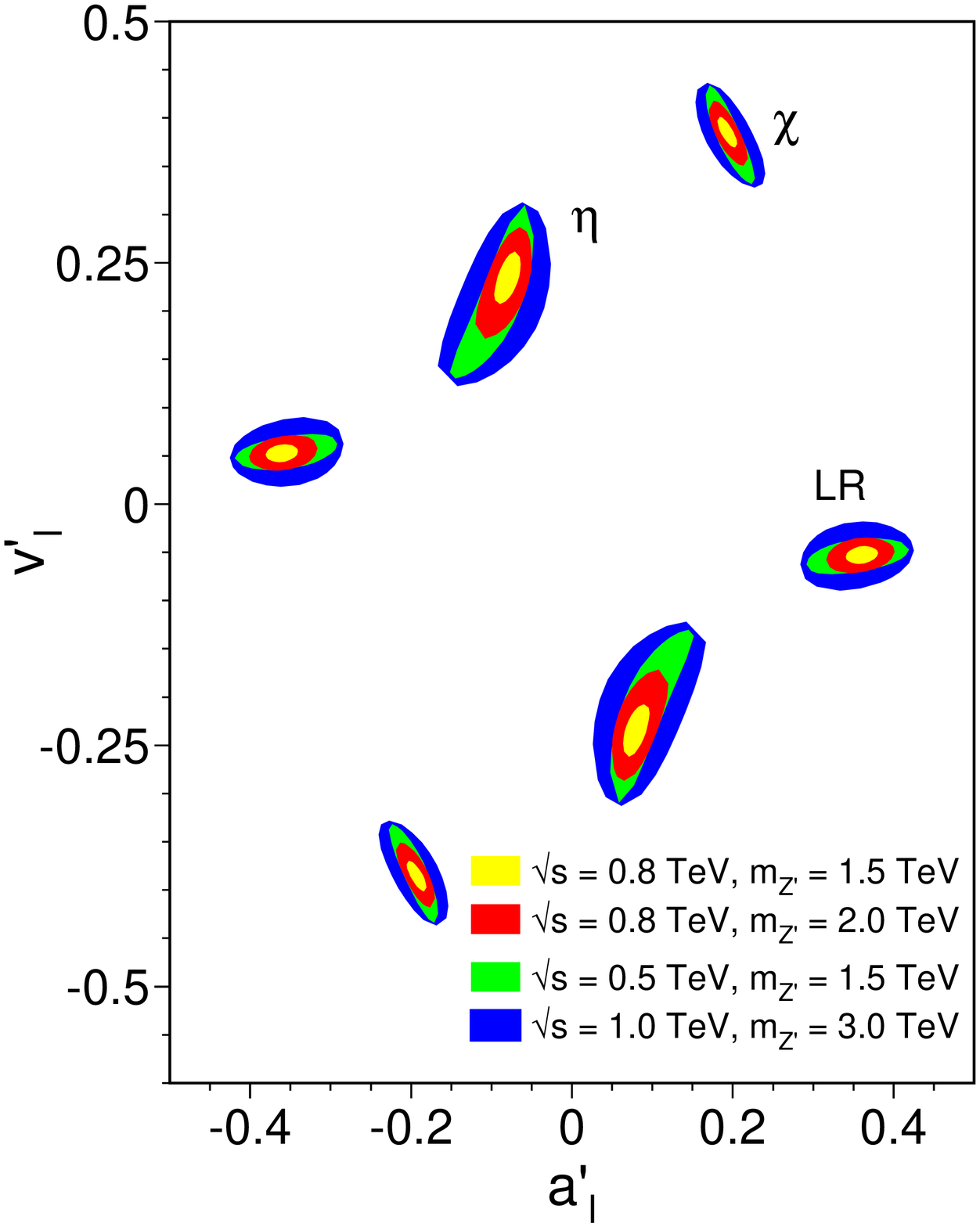}
\caption{Z$'$ masses and couplings in extended SO(10) and E(6) gauge theories; Ref.~\cite{r39}.}
\label{F16}
\end{center}
\end{figure}
%%%%%%%%%%%%%%%%%

Such intermediate gauge bosons can be searched for at LHC for masses
up to about 5 TeV. The role of ILC is twofold. First, by analyzing the
effect of virtual $Z'$ s-channel exchange on the cross sections and
angular distributions of fermion pair production, $e^+e^- \to f
\bar{f}$, the sensitivity to new gauge boson scales can be extended
significantly, cf.\ Fig.~\ref{F16}, in SO(10) LR symmetric theories up
to $\simeq 15$ TeV at ILC (and up to $\simeq 35$ TeV at
CLIC~\cite{r39.1}). Second, the couplings of the new $Z'$ boson to SM
fermions can be determined very precisely, Fig.~\ref{F16}. The various
models can obviously be discriminated quite clearly and the nature of
the underlying gauge symmetry can be identified.

\section{EXTRA SPACE DIMENSIONS}

A large variety of models have been developed in which the ordinary
4-dimensional space-time is extended to higher dimensions already at
energies of order 1 TeV.  The ILC potential in analyzing such models,
in which the extra dimensions are compactified at low scales, will be
illustrated in two examples.  

\vspace{\baselineskip}
\noindent\underline{\it ADD Scenario}    

\vspace{0.5\baselineskip}
In the ADD scenario \cite{r1.8} gravity extends from the brane 
on which the fields of the Standard
Model are located, to the higher ${\rm{D}}= 4+\delta$ dimensions. It
becomes strong in the extended space already at the fundamental Planck
scale $\Lambda_{\rm{D}}$ of order TeV, much below the effective
standard Planck scale $\Lambda_{\rm{Pl}}$ of order $10^{19}$ GeV, and
it appears weak only if projected onto the 4-dimensional SM brane. The
radii of the compactified higher dimensions are related to the Planck
scale by $\Lambda_{\rm{Pl}}^2 = R^\delta
\Lambda_{\rm{D}}^{2+\delta}$. The associated Kaluza-Klein states with
masses $\sim n/R$ densely populate a tower with energy spacings of a
small fraction of eV up to a few MeV, depending on the number of extra
space dimensions.  

\begin{figure}[h]
\begin{center}
\includegraphics[height=6cm]{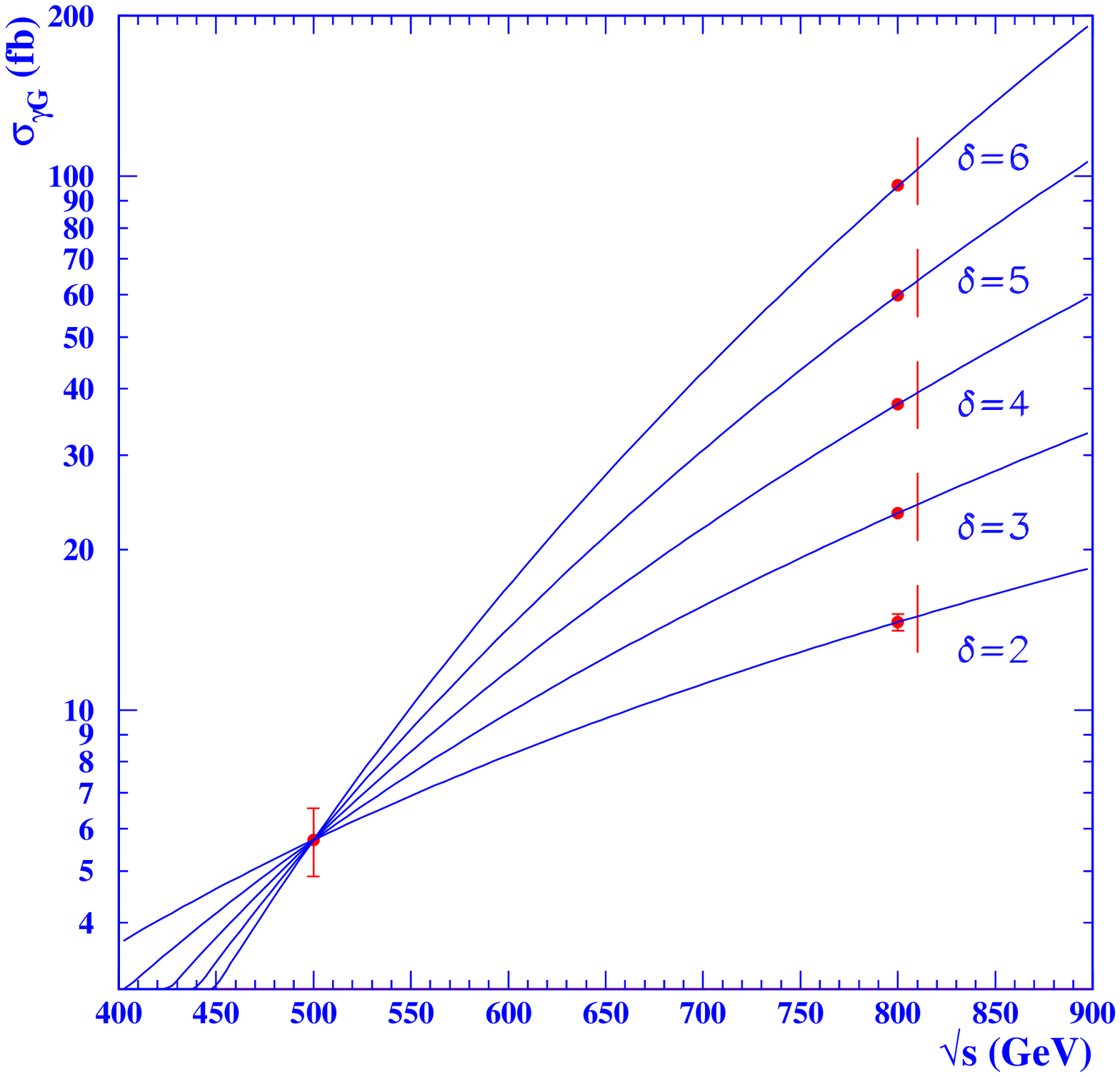}
\raisebox{-5mm}{
\includegraphics[height=9.4cm,angle=90]{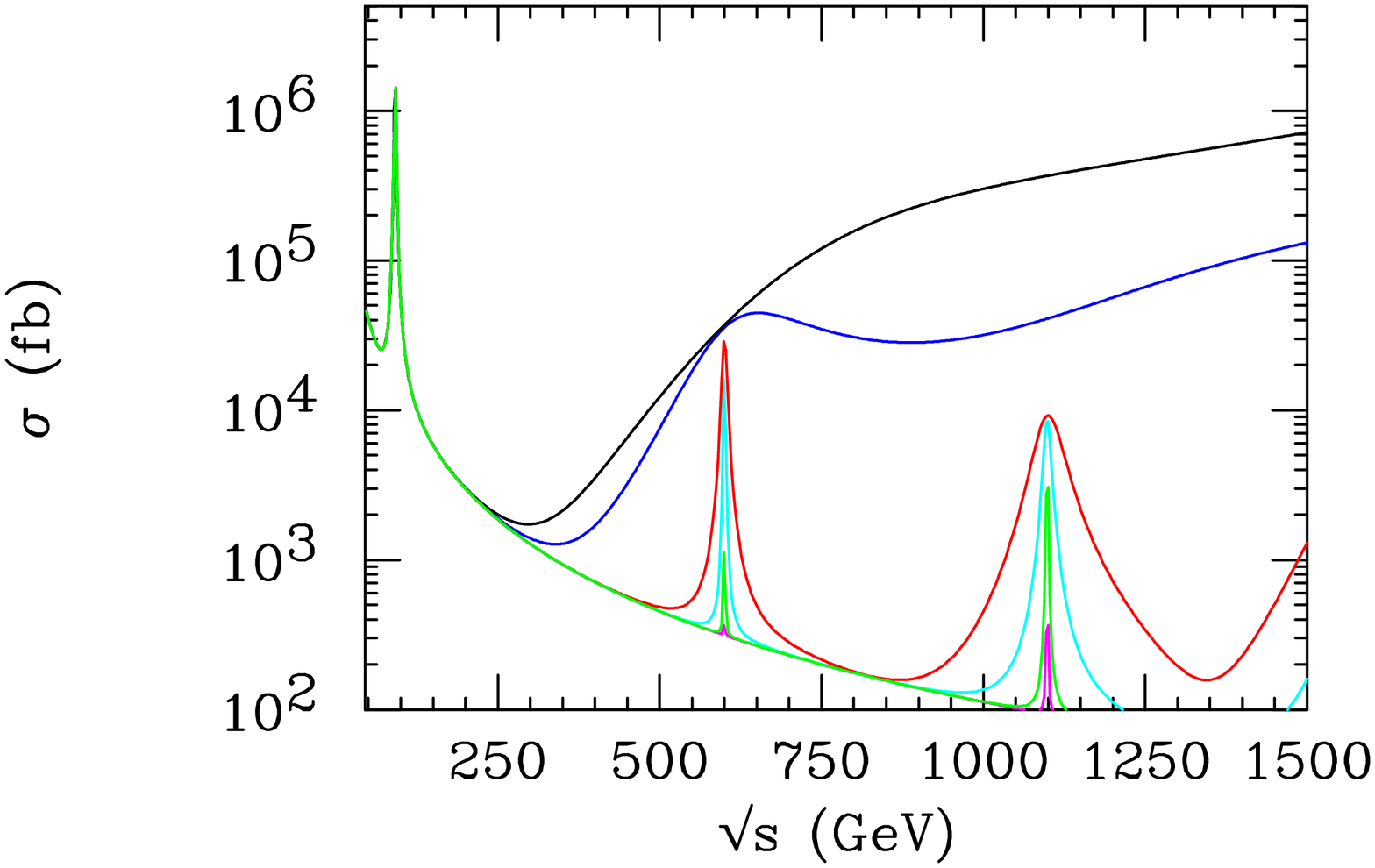}}
\caption{Left: Measuring the Planck scale and the number of dimensions in ADD scenarios; 
         Ref.~\cite{r40.2}. 
         Right: Kaluza-Klein excitations in RS scenarios for various values of the curvature
         $k$; Ref.~\cite{r41}.}
\label{F17}
\end{center}
\end{figure}

At $e^+e^-$ linear colliders the two crucial parameters of the ADD
model, the fundamental Planck scale $\Lambda_{\rm{D}}$ and the number
$\delta$ of extra space dimensions, can be measured by varying the cm
energy of the collider. The cross section for the process of single
$\gamma$ production,
\begin{equation}
e^+ e^- \to \gamma + G_{\rm KK} 
\end{equation}
where $G_{\rm KK}$ denotes the sum over the invisible graviton states
of the Kaluza-Klein tower, depends on these two parameters in the form
\cite{r40.1}
\begin{equation}
\sigma(e^+e^- \to \gamma + \Eslash) 
  = \frac{c_\delta}{\Lambda^2_{\rm D}}
  \left(\frac{\sqrt{s}}{\Lambda_{\rm{D}}}\right)^\delta \, . 
\end{equation}
Thus, the larger the number of extra dimensions the stronger would be
the rise of the cross section for single isolated photons with the
collider energy, Fig.~\ref{F17}.  

\vspace{\baselineskip}
\noindent\underline{\it RS Scenario}  

\vspace{0.5\baselineskip} 
While in the previous model space is flat in the extra dimensions, it
is curved in the RS model \cite{r1.9}. The curvature in the extra
fifth dimension is described by an exponential warp factor $\exp(- 2 k
r_c \phi)$, characterized by the compactification radius $r_c$ and the
curvature $k$. The coordinate $\phi$ spans the distance between the
gravity brane located at $\phi = 0$ and the SM brane located at $\phi
= \pi$. Since the scale of physical processes on the SM brane is given
by $\Lambda_{\rm{SM}} = \Lambda_{\rm{Pl}} \exp(-k r_c \pi) \sim$ 1
TeV, the compactification radius $r_c$ is estimated to be, roughly,
one order of magnitude larger than the curvature radius $k^{-1}$,
while $k$ itself is of the order of the effective 4-dimensional Planck
scale. The characteristics of our eigen-world on the 4-dimensional SM
brane are described by the two parameters $k$ and $r_c$, with the
second parameter generally substituted by $\Lambda_{\rm SM}$.
  
The Kaluza-Klein tower of the gravitons on the SM brane is
characteristically different from towers associated with flat spaces,
the sequence of masses \cite{r41} given by
\begin{equation}
M_n = x_n \,k\, \exp(-k r_c \pi) = x_n \Lambda_{\rm{SM}} \,k\, / \Lambda_{\rm{Pl}} 
\end{equation}
where $x_n$ are the roots of the first-order Bessel functions. Such
states can be searched for in fermion pair production $e^+e^- \to
\mu^+ \mu^-$, affecting this process by resonant s-channel exchanges. Fixing 
the lowest KK state to a mass of 600 GeV, the sequence of KK excitations
is displayed in Fig.~\ref{F17}. The width of
%%%%%%%%%%%%%%%%%
%%%%%%%%%%%%%%%%%
the KK states depends on the curvature $k \sim \Lambda_{\rm{Pl}}$
in the fifth dimension.  The cross sections turn out to be very large
if the parameters are such that the lowest KK states can be generated
at the collider as an s-channel resonance.

\section{COSMOLOGY CONNECTION}

Collider physics programs focus in connection with cosmology presently
on two fundamental problems, cf.\ Ref.~\cite{r42}:
\begin{itemize}
\item[--]
  the mechanism responsible for the baryon asymmetry:
  \hfill
  $\rho_{\rm B} = \parbox{2cm}{$4.0 \pm 0.4 \%$}$\mbox{\hspace*{5cm}}
\item[--]
  the particle character of cold dark matter: 
  \hfill
  $\rho_{\rm CDM} = \parbox{2cm}{$23 \pm 4 \%$}$\mbox{\hspace*{5cm}}
\end{itemize}
These central problems of physics cannot be solved within the
framework of the Standard Model. Various solutions have been worked
out which require experiments at high energy colliders to establish
the proposed mechanism for generating the baryon asymmetry in the
universe and for clarifying the nature of cold dark matter.  Even if a
single particle species were the main component of cold
dark matter in the universe, the theoretical origin will in general be
so complex that laboratory experiments are required to achieve the
proper understanding of this phenomenon.

\subsection{Baryon Asymmetry}

Two approaches for generating the baryon asymmetry are widely
discussed in the literature: baryogenesis mediated by leptogenesis, and
electroweak baryogenesis based on the supersymmetric extension of the
Standard Model.  

\vspace{\baselineskip}
\noindent\underline{\it Leptogenesis}  

\vspace{0.5\baselineskip}
If leptogenesis \cite{r43.1} is the origin of the observed baryon asymmetry, 
the roots of this phenomenon are located near the Planck
scale. CP-violating decays of heavy right-handed Majorana neutrinos
generate a lepton asymmetry which is transferred to the quark/baryon
sector by sphaleron processes. Heavy neutrino mass scales as
introduced in the seesaw mechanism for generating light neutrino
masses and the size of the light neutrino masses needed for
leptogenesis define a self-consistent frame which is compatible with
all experimental observations \cite{r43.2}.  

As shown in the preceding chapter, in some SUSY models the size of the
heavy seesaw scales can be related to the values of the charged and
neutral slepton masses \cite{r38.2}. A sum rule relates the difference
between the slepton masses of the first two and the third generation
to the mass of the heavy right-handed Majorana neutrino in the third
generation within SO(10) based supergravity theories. In this way the
size of the seesaw scale can well be estimated.

%\newpage
\vspace{\baselineskip}
\noindent\underline{\it Electroweak Baryogenesis in Supersymmetry}  

\vspace{0.5\baselineskip}
One of Sakharov's conditions for generating the baryon asymmetry of
the universe requires a departure from thermal equilibrium. If
triggered by sphaleron processes at the electroweak phase transition,
the transition must be sufficiently strong of first order. Given the
present bounds on the Higgs mass, this cannot be realized in the
Standard Model. However, since top and stop fields modify the Higgs
potential strongly through radiative corrections, supersymmetry
scenarios can give rise to first-order transitions, cf. Ref.~\cite{r44.1}. 
The parameter space of the MSSM is tightly constrained in this case: 
The mass of the light Higgs boson is bounded by 120 GeV from above, 
and the mass of the light stop quark is required to be smaller 
than the top quark mass, cf. Ref.~\cite{r44.2}.  

This scenario suggests that the mass of the stop quark is only
slightly larger than the lightest neutralino (LSP) mass. The correct
density of cold dark matter is generated by stop-neutralino coannihilation 
in this region of parameter space, leading to tight constraints for the masses
of the two particles.

While studies of the light stop quark are very difficult at hadron
colliders if the main decay channel is the two-body decay $\tilde{t}_1
\to c \tilde{\chi}_1^0$ with a low-energy charm jet in the final
state, the clean environment of an $e^+e^-$ collider
%%%%%%%%%%%%%%%%%
\begin{figure}[h]
\begin{center}
\includegraphics[height=7cm]{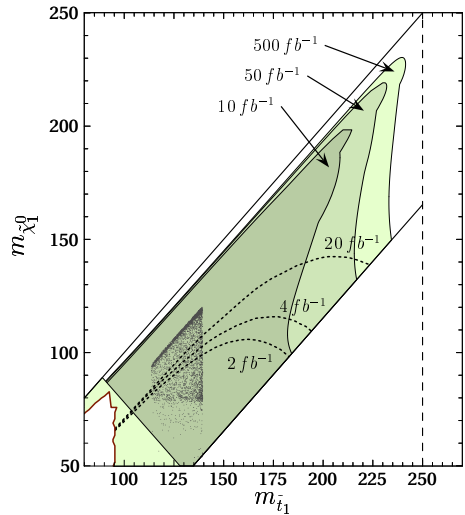}
\caption{ILC coverage of stop/neutralino parameter space, dark grey points, in the MSSM 
         based electroweak phase transition; Ref.~\cite{r44.2}.}
\label{F18}
\end{center}
\end{figure}
%%%%%%%%%%%%%%%%%
allows for precision studies of the system also in such
configurations. This is demonstrated in Fig.~\ref{F18} which proves
that a linear collider covers completely the region of dark gray
points which are compatible with precision measurements of the cold
dark matter density.

\subsection{Cold Dark Matter}
 
Cold dark matter (CDM) is the dominant component of matter in the
universe. So far it has not been possible yet to reveal its
microscopic nature. Attempts to solve this problem form an intimate
link between cosmology and particle physics. CDM may be a complex
structure and a mixture of several components. Theoretical particle
physics offers hypothetical particle candidates which could be
discovered, if realized in nature, in the next generation of
accelerators. After determining the properties of candidate particles
in laboratory experiments, their density in the universe can be
predicted and the prediction can be confronted with cosmological
precision measurements. In addition, compatibility with direct and
indirect search experiments must be checked.  In this way a closed
circle may evolve which provides a self-consistent picture of the
nature of cold dark matter and its distribution in the universe.

Theories which provide a CDM candidate must have a conserved parity
quantum number.  Examples
are R parity in supersymmetric models, KK parity in extra-dimensional
models, or T~parity in Little-Higgs theories.  The lightest particle
with odd parity is then stable, must be charge- and color-neutral, and
thus provides a CDM candidate.  If this particle is in or below the TeV
mass range and interacts with matter, it will be seen via
missing-energy signatures at LHC.  At ILC, a precise determination
of its mass and interactions is possible due to kinematical
hermeticity and low background, independently of the embedding
theory. 

Among the candidate theories, two specific examples will be summarized
briefly to illuminate the ILC potential in clarifying the nature of
cold dark matter particles.  The examples chosen are the 
supersymmetric extension of the Standard Model embedded in minimal
supergravity (constrained MSSM) in which the lightest neutralino is
the cold dark matter particle, and a supergravity theory in which the
gravitino is identified with this particle. In the first example, the
characteristic parameters are the gravitino mass, with a value near
the electroweak scale, and the weak interactions of CDM. In the second
example, CDM interacts only through gravity.

\vspace{\baselineskip}
\noindent\underline{\it Neutralino Cold Dark Matter}  

\begin{figure}[h]
\begin{center}
\raisebox{-2mm}{
\includegraphics[height=5.9cm]{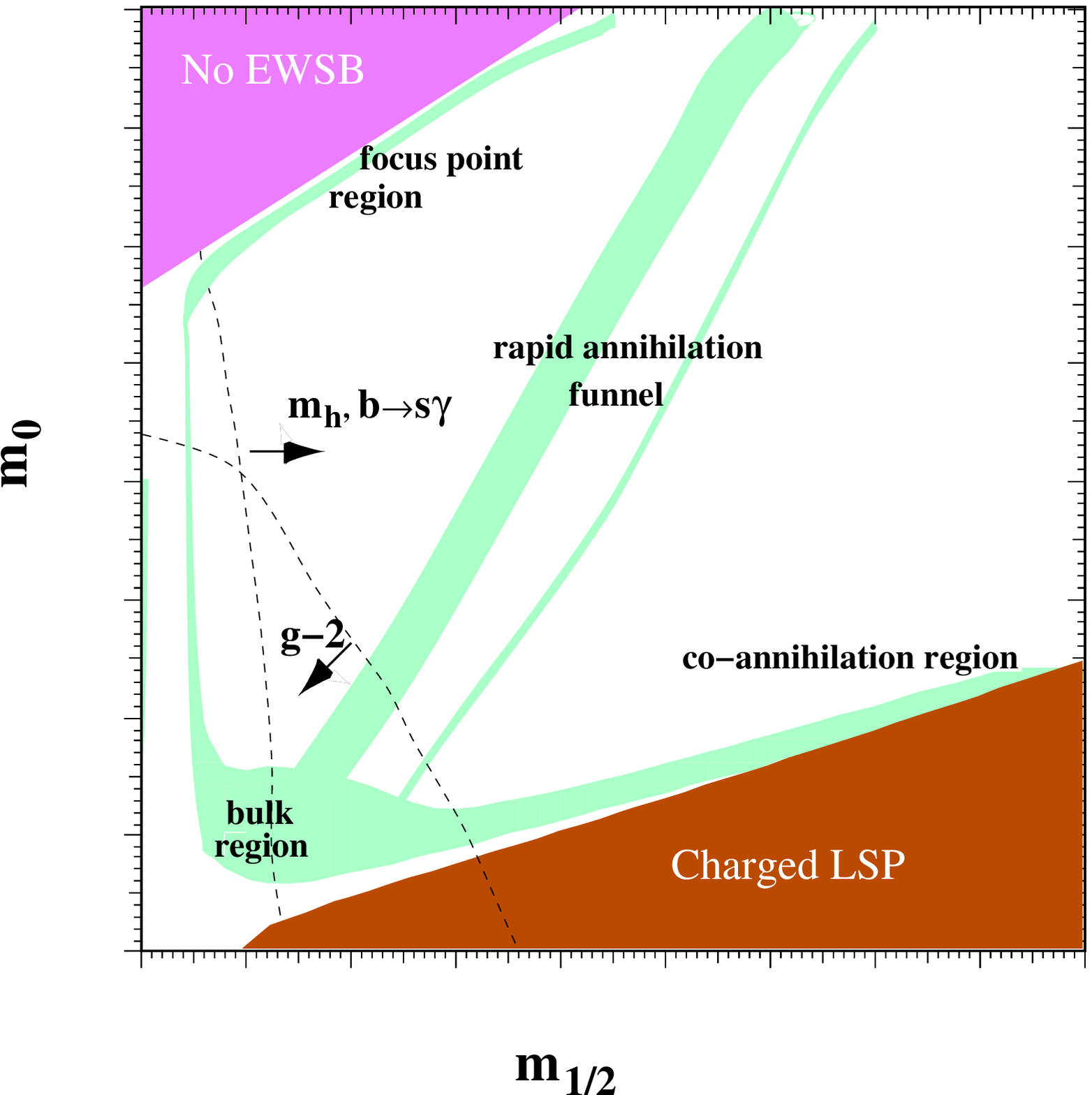}} \hskip 5mm
\includegraphics[height=6.3cm]{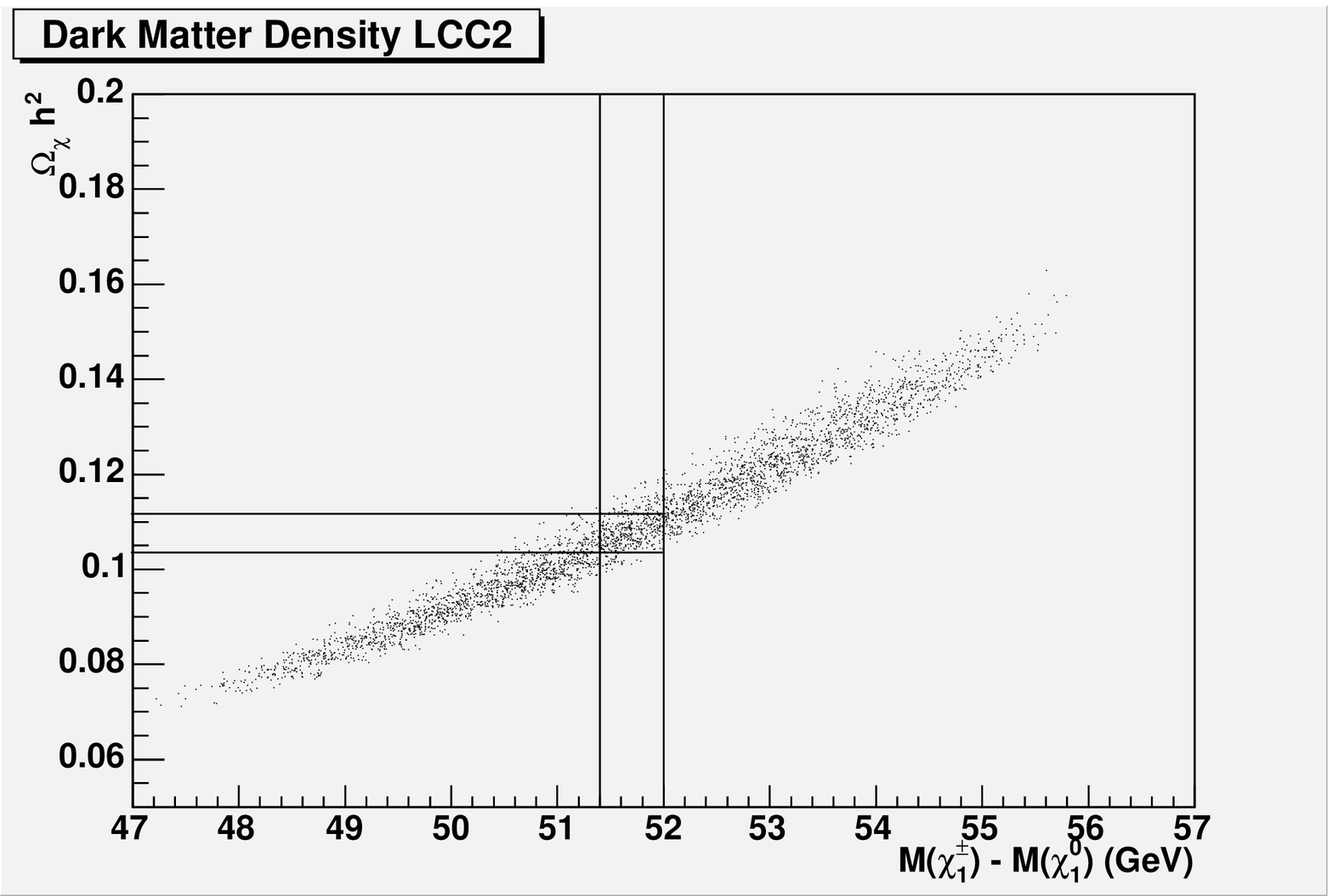}
\caption{Left: Characteristics of mSUGRA parameter regions constrained
         by precision data on the relic density; cf.
         Ref.~\cite{r42}. Right: Sensitivity of the prediction for the
         relic density from parameter measurements in the focus point
         region; Ref.~\cite{r45.2}, see also Refs.~\cite{r45.3}.}
\label{F19}
\end{center}
\end{figure}

\vspace{0.5\baselineskip}
In the mSUGRA parameter range four characteristic areas have been
identified in which the observed relic density~\cite{r45.1}
can be accommodated, cf.\ Fig.~\ref{F19}, and they have recently 
been studied systematically \cite{r45.2,r45.3,r45.3.1}.      
%%%%%%%%%%%%%%%%%
%%%%%%%%%%%%%%%%% 

{\it (i)} In the {\it bulk region} the gaugino mass parameter $M_{1/2}$ and
the scalar mass parameter $M_0$ are both in the area surrounding the
electroweak scale. Neutralino pairs annihilate into fermion
pairs. This area, including the benchmark point SPS1a$'$, has been
studied very thoroughly for LHC as well as ILC experiments. From the
ILC studies the CDM density is expected to be determined within an
accuracy of about 2\%, thus matching the precision expected from
measurements of the Planck satellite in the near future.  

{\it (ii)} In the {\it focus point region} the gaugino mass parameter
remains moderate but the scalar mass parameter is very large. While
the spectrum of charginos and neutralinos appears accessible at ILC,
sleptons can be produced, if at all, only at a multi-TeV collider.  Split
Supersymmetry, where sleptons are completely inaccessible, is an
extreme case of this scenario.  Neutralino
pairs annihilate primarily to gauge bosons. The prediction of the
relic density is strongly correlated with the mass difference between
the lightest chargino and neutralino.  At the specific point studied
in detail, cf.\ Fig.~\ref{F19}, the relic density is predicted 
\cite{r45.2} at a level of 3.5\%.  

{\it (iii)} The $\tilde{\tau} \tilde{\chi}$ {\it coannihilation region} 
with moderate to large $M_{1/2}$ and moderate $M_0$ is difficult to
explore experimentally as $\tilde{\tau} \to \tau \tilde{\chi}_1^0$
decays must be studied in which stau and neutralino are close in mass
so that the visible $\tau$ in the final state carries only a small
amount of energy and is hard to detect.  

{\it (iv)} In the {\it funnel region} neutralino annihilation is
mediated by an s-channel Higgs boson.  This scenario has received some
attention in connection with an astrophysical observation: Galactic
photon spectra measured with the EGRET instrument appear to be in
excess over the conventionally expected yield. Neutral pions emerging
from neutralino annihilation to $b$-jet pairs have been proposed as an
explanation of the excess \cite{r45.4}. Neutralino masses in the range
between 50 and 100 GeV and sfermion masses around 1 TeV would be
predicted this context. The size and the shape of the excess however
depend on the modeling of conventional background sources while
exclusive neutralino annihilation into a pair of photons could provide
a peak in the 2-photon energy that would be a model-independent
observation of the annihilation of two CDM particles.

\vspace{\baselineskip}
\noindent\underline{\it Gravitino Cold Dark Matter}  

\vspace{0.5\baselineskip}
In supergravity models the gravitino $\tilde{G}$ itself may be the
lightest supersymmetric particle, building up the dominant CDM component,
cf. Ref.~\cite{r45.5}. 
In such a scenario, with a gravitino mass in the range of 100 GeV
[in contrast to gauge-mediated supersymmetry breaking with
very light gravitino mass], the lifetime of the
next-to-lightest supersymmetric particle can become macroscopic as the
gravitino coupling is only of gravitational strength. 
The lifetime of the NLSP $\tilde{\tau}$,
\begin{equation}
\tau [\tilde{\tau} \to \tau + \tilde{G}] = const \, \times \, M^2_{\tilde{G}}
                                           M^2_{\rm Pl} / M^5_{\tilde{\tau}} 
\end{equation}
can extend up to several months, suggesting special experimental
efforts to catch the long-lived $\tilde{\tau}$'s and to measure their
lifetime \cite{r46}. Production in $e^+e^-$ annihilation determines the
$\tilde{\tau}$ mass, the observation of the $\tau$ energy in the
$\tilde{\tau}$ decay the gravitino mass. The measurement of the
lifetime can subsequently be exploited to determine the Planck scale
$M_{\rm Pl}$, a unique opportunity in a laboratory experiment.

\section{SUMMARY}
 
The ILC can contribute to solutions of key questions in physics,
\begin{itemize}
\item[--]
  {\it Electroweak Symmetry Breaking}: The Higgs mechanism {\it sui
    generis} can be established for breaking the electroweak symmetries 
  and generating the masses of the fundamental particles.  
\item[--]
  {\it Grand and Ultimate Unification}: A comprehensive and
  high-resolution picture of supersymmetry can be drawn by coherent
  analyses of hadron and lepton collider experiments. Thus the colliders
  may become telescopes to the physics scenario near the Planck scale
  where particle physics is linked with gravity and where the basic
  roots of physics are expected to be located.  
\item[--]
  {\it Extra Space Dimensions}: The parameters of an extended
  space-time picture can be determined, the fundamental Planck scale and
  the number of extra dimensions.  New Kaluza-Klein states can either be
  generated directly or their effect on Standard Model processes can be
  explored.  
\item[--]
  {\it Cosmology Connection}: Drawing a microscopic picture of particles
  building up cold dark matter, the basis necessary for the
  understanding of matter in the universe can be provided by collider
  experiments. In addition, crucial elements for explaining the baryon
  asymmetry in the universe can be reconstructed.  
\end{itemize}
Collider experiments will thus be essential instruments for unraveling
the fundamental laws of nature.
  
%%%%%%%%%%%%%%%%%%%%%%%%%%%%%%%%%%%%%%%%%%%%%%%%%%%%%%%%%%%%%%%%%%%%%%%%%%%%%%%%%%%%%%%%%%%

%%%%%%%%%%%%%%%%%%%%%%%%%%%%%%%%%%%%%%%%%%%%%%%%%%%%%%%%%%%%%%%%%%%%%%%%%%%%%%%%%%%%%%%%%%%
\end{document}